\documentclass[12pt]{article}
\usepackage[dvips]{graphicx}
\usepackage{textcomp}
\usepackage{amssymb}

\newcommand{\mc}{\multicolumn}

\begin{document}

\begin{titlepage}
\vskip0.5cm
\begin{flushright}
\end{flushright}
\vskip0.5cm
\begin{center}
{\Large\bf The specific heat of thin films near the $\lambda$-transition:  
 A Monte Carlo study of an improved three-dimensional 
 lattice model} 
\end{center}

\centerline{
Martin Hasenbusch
}
\vskip 0.3cm
\centerline{\sl  Institut f\"ur Physik, Humboldt-Universit\"at zu Berlin
}
\centerline{\sl  Newtonstr. 15, 12489 Berlin, Germany}
\vskip 0.1cm
\centerline{\sl
e--mail: \hskip 0.1cm
 Martin.Hasenbusch@physik.hu-berlin.de}
\vskip 0.4cm
\begin{abstract}
We study the finite size scaling behaviour of the specific heat of thin 
films in the 
neighbourhood of the $\lambda$-transition. To this end we have simulated 
the improved two-component $\phi^4$ model on the simple cubic lattice.
We employ free boundary conditions in the short direction to mimic the 
vanishing order parameter at the boundaries of a $^4$He film. Most of 
our simulations are performed for the thicknesses $L_0=8$, $16$ and $32$
of the film.
It turns out that one has to take into account corrections  $\propto L_0^{-1}$
to obtain a good collapse of the finite size scaling functions obtained 
from different $L_0$.
Our results are compared with those obtained from experiments on thin films
of $^4$He near the $\lambda$-transition, from field theory and from previous
Monte Carlo simulations.
\end{abstract}
\end{titlepage}

\section{Introduction}
In the neighbourhood of a second order phase transition the behaviour of 
various  quantities is governed by power laws.
For example the correlation length diverges as
\begin{equation}
\label{xipower}
\xi \simeq \xi_{0,\pm} |t|^{-\nu} \;,
\end{equation}
where $t=(T-T_c)/T_c$ is the reduced temperature, $\xi_{0,+}$ and $\xi_{0,-}$ 
are the amplitudes in the high and the low temperature phase, respectively,
and $\nu$ is the critical exponent of the correlation length. $T_c$ is the 
critical temperature, where the phase transition occurs. \footnote{In the case 
of the $\lambda$-transition of $^4$He we shall denote the critical
temperature by $T_{\lambda}$.}
The specific heat behaves as
\begin{equation}
\label{cpower}
C \simeq A_{\pm} |t|^{-\alpha} + B  \;,
\end{equation}
where $A_{+}$ and $A_{-}$ are the amplitudes
in the high and the low temperature phase, respectively, and $B$ is an 
analytic background,
which has to be taken into account here, since the critical exponent $\alpha$
of the specific heat is negative for the three-dimensional XY universality
class. A universality class is characterised by the dimension of the system, 
the range of the interaction and the symmetry of the order parameter.
Critical phenomena 
can be understood in the framework of the  Renormalization Group (RG).
For reviews on critical phenomena and the Renormalization Group see e.g.
\cite{WiKo,Fisher74,Fisher98,PeVi02}.
The XY universality class in three dimensions with short range 
interactions is of particular interest, since the 
$\lambda$-transition of $^4$He shares this universality class. 
Results for critical exponents and amplitude ratios such as $A_+/A_-$ obtained
at the $\lambda$-transition, are more precise than those from
experiments on other systems.

It is an interesting question how the critical behaviour is modified by
a confining geometry. If the system is finite in all directions, 
thermodynamic functions have to be analytic functions. I.e. a singular
behaviour like eqs.~(\ref{xipower},\ref{cpower}) is excluded. 
As a remnant of such singularities there remains a peak in the
neighbourhood of the transition.  With increasing linear extension
the hight of the peak increases and the temperature of the maximum approaches
the critical temperature.
This  behaviour is described by the theory of finite size scaling (FSS). 
For reviews see \cite{Barber,Privman}.

In the present work we study thin films. Thin films are finite in one direction
and infinite in the other two directions. In this case singular behaviour is 
still possible. However the associated phase transition belongs to the 
two-dimensional universality class. I.e. in the case of $U(1)$ symmetry,
a Kosterlitz-Thouless (KT) transition \cite{KT,Jo77,AmGoGr80} is expected.
In our recent work \cite{myKTfilm} we have focused on the study 
of this transition and the scaling of the transition temperature with the
thickness of the film.

Here we investigate the behaviour of the specific heat of thin films 
in the neighbourhood of the $\lambda$-transition. The specific heat has
been studied in a number of experiments on thin films of fluid $^4$He and 
$^3$He-$^4$He mixtures near the $\lambda$-transition.
For recent reviews see \cite{BaHaLiDu07,GaKiMoDi08}.

The physics of thin films is governed by the ratio $L_0/\xi$, where 
$\xi$ is a correlation length of the bulk system and $L_0$ the thickness
of the thin film. For $L_0 \gg \xi$ the behaviour of the film is essentially
given by the thermodynamic limit of the three-dimensional system. In 
the critical region, when $\xi$ gets close to $L_0$ or even larger,
the behaviour deviates from the three-dimensional one and is characterised 
by universal functions of $L_0/\xi$.

In particular, the behaviour of the specific heat can be  described
by the universal scaling function
\begin{equation}
\label{sc1a}
[C_{bulk}(t) - C(t,L_0)] A^{-1} |t|^{\alpha}=
\tilde g_{2,R}(L_0/\xi) =  g_{2,R}(t [L_{0}/\xi_0]^{1/\nu})
\end{equation}
where $C_{bulk}(t)$ is the specific heat of the three-dimensional 
thermodynamic limit and $C(t,L_0)$ the specific heat of a film of 
thickness $L_0$. As $A$ either $A_+$ or $A_-$ can be taken and analogously
as $\xi_0$ either $\xi_{0,+}$ or $\xi_{0,-}$. 
In the last step of the equation we have used 
eq.~(\ref{xipower}). Following RG-theory, the analytic background $B$ 
is the same for the bulk and the thin film. Therefore it cancels
in the difference that is considered here. Alternatively
one might consider
\begin{equation}
\label{sc1b}
[C_{bulk}(t_0) - C(t,L_0)] A^{-1} |t|^{\alpha}=
\tilde g_{1,R}(L_0/\xi) =  g_{1,R}(t [L_{0}/\xi_0]^{1/\nu})
\end{equation}
where $t_0$ is chosen such that $\xi(t_0)=L_0$ in the high temperature
phase.
Multiplying  eqs.~(\ref{sc1a},\ref{sc1b}) by 
$(|t| [L_0/\xi_0]^{1/\nu})^{-\alpha}$
one arrives at
\begin{equation}
\label{scaling2L0}
[C_{bulk}(t) - C(t,L_0)] A^{-1} \xi_0^{\alpha/\nu} L_0^{-\alpha/\nu}=
\tilde f_{2,R}(L_0/\xi) =  f_{2,R}(t [L_{0}/\xi_0]^{1/\nu})
\end{equation}
and
\begin{equation}
\label{scaling1}
[C_{bulk}(t_0) - C(t,L_0)] A^{-1} \xi_0^{\alpha/\nu}  L_0^{-\alpha/\nu}=
\tilde f_{1,R}(L_0/\xi) =  f_{1,R}(t [L_{0}/\xi_0]^{1/\nu})
\end{equation}
respectively. Often in the literature the factors $A^{-1}$ and 
$A^{-1} \xi_0^{\alpha/\nu}$ are omitted and
$t L_{0}^{1/\nu}$ is used as argument of the scaling function. 
This 
poses no problem as long as films of different thicknesses of the 
same system are considered. However, comparing films of e.g. $^4$He
at different pressures or $^3$He-$^4$He mixtures at different concentrations
of $^3$He this fact has to be taken into account. The same holds for 
the comparison of such experimental results with those obtained from 
lattice models or field theory. In the following we shall use the notation
$f_1=A \xi_0^{-\alpha/\nu} f_{1,R}$ and $f_2=A \xi_0^{-\alpha/\nu} f_{2,R}$.

These universal scaling functions have been determined by a number of 
experiments on $^4$He and mixtures of $^3$He and $^4$He. 
In the high temperature phase, the data follow nicely the prediction
of finite size scaling as can be seen e.g. from figure 14 of  
\cite{GaKiMoDi08}. In this figure, 
$[C_{bulk}(t) - C(t,L_0)] |t|^{\alpha}$ is plotted as a function of
$|t| [L_0/\xi_0]^{1/\nu}$ for films of  $^4$He at vapour pressure 
of thicknesses 483 $\AA$ up to $57  \mu$m. The data for different 
thicknesses fall nicely on top of each other.
In their figure 20 the authors of  \cite{GaKiMoDi08}  have plotted 
data for the low temperature phase in an analogous way. Up to 
$|t| L_0^{1/\nu} \approx 5$ the data for different thicknesses fall nicely 
on top of each other. However for larger values of $|t| L_0^{1/\nu}$ the 
data start to scatter. This is most pronounced at 
$|t| L_0^{1/\nu} \approx 10$, where the function assumes a minimum. There
is a factor of about $3.5$ between the value for the thinnest and the 
value for the thickest film.
From $|t| L_0^{1/\nu} \approx 20$ up to $\approx 100$ there is a factor
of about two between the thinnest and the thickest of the films. Note that
in  figure 20 of  \cite{GaKiMoDi08} $L_0$ is given in $\AA$ and 
$t=1-T/T_{\lambda}$.

In the case of superfluid helium the order parameter is a complex wavefunction.
This wave function vanishes at the boundaries of the film. 
In order to mimic this in theoretical models, Dirichlet boundary 
conditions with vanishing field are employed.

Using such boundary conditions, the scaling function $f_2$ has been
calculated by using the $\epsilon$-expansion to 
O$(\epsilon)$ \cite{KrDi92,KrDi92b}. 
The coefficients of O$(1)$ and O$(\epsilon)$ are numerically of similar size.
Therefore one should not expect quantitatively accurate results for $f_2$
obtained this way.
Both $f_1$ and $f_2$ have been computed by using perturbation theory
in three dimensions fixed \cite{ScWaDoFr90,Dohm93,SuDo94} in one-loop
approximation. In the high temperature phase and at the 
critical point of the bulk system, the experimental results
are fairly well reproduced.
However in the low temperature phase, in particular  close
to the KT transition, no accurate predictions can be obtained.

Also Monte Carlo simulations of the standard XY model on a simple cubic 
lattice have been performed to determine the specific heat of thin 
films. For a precise definition of the XY model see below.
In \cite{ScMa95} staggered boundary conditions have been used to obtain 
a vanishing order parameter at the boundaries. The authors of \cite{NhMa03}
have employed free (in their notation ``open'') boundary conditions as 
we do in the present work. In both cases the authors have computed 
the finite size scaling function $f_1$.  In \cite{ScMa95} the authors have 
simulated lattices of a thickness up to $L_0=24$ lattice units,
while in \cite{NhMa03} the thicknesses $L_0=12$, $14$ and $16$ have 
been studied.  The results of \cite{ScMa95} and \cite{NhMa03} for $f_1$
agree. There is also a reasonable match with experiments on helium
films.

The purpose of the present paper is to compute the finite size 
scaling function $f_2$
for the first time by using Monte Carlo simulations of a lattice model. 
Furthermore we carefully study corrections to scaling, allowing us to 
quantify the error of our result for the  finite size scaling function.

For finite systems we expect that the leading corrections are
$\propto L_0^{-\omega}$, irrespective of the type of the boundary
conditions \cite{Barber}. 
The numerical value of the correction exponent is $\omega=0.785(20)$ 
for the XY universality class in three dimensions \cite{recentXY}; similar
results are obtained with field-theoretic methods; see e.g. \cite{PeVi02}.
In order to avoid these corrections, we study an improved model. 
In improved models the amplitude of corrections $\propto L_0^{-\omega}$ 
vanishes or in practise, it is so  small that its effect can be ignored.
The precise definition of the model that we have simulated is given below.

On top of the restricted geometry, free boundary conditions introduce 
new physical effects. For a discussion see e.g. reviews on surface critical
phenomena \cite{Binder,Diehl86}. In fact,
free boundary conditions lead to additional corrections to scaling.
The leading one is $\propto L_0^{-1}$  \cite{DiDiEi83}; 
it can be cast in the form $L_{0,eff} = L_0 + L_s$. 
\footnote{
In the literature, replacing $L_0$ by $L_{0,eff} =L_0 + L_s$ to account for
surface corrections, was first discussed by Capehart and Fisher \cite{CaFi76}
in the context of the surface susceptibility of Ising films.}
In \cite{myKTfilm} we have obtained the accurate numerical estimate 
$L_s=1.02(7)$ for the model that we simulate here.

This paper is organized as follows:  In the next section we define the 
lattice model that we have simulated and the observables that we have
computed. In section three we discuss how corrections caused by the 
free boundary conditions affect the finite size scaling behaviour of 
the specific heat. Next we discuss the details of our simulations.
Based on these simulations we compute the scaling functions $f_1$ and $f_2$.
We compare our results with those from experiments on thin films
of $^4$He, field theoretic methods and previous Monte Carlo simulations.

\section{The model and the observables}
\subsection{The two component $\phi^4$ model}
We study the two component $\phi^4$ model on a simple cubic lattice.
We  label the sites of the lattice by
$x=(x_0,x_1,x_2)$. The components of $x$ might assume the values 
$x_i \in \{1,2,...,L_i\}$.  We simulate lattices
of the size $L_1=L_2=L$ and $L_0 \ll L$.  In 1 and 2-direction we employ
periodic boundary conditions and free boundary conditions in 0-direction.
This means that the sites with $x_0=1$ and $x_0=L_0$ have only five nearest
neighbours.
This type of boundary conditions could be interpreted as Dirichlet 
boundary conditions with $0$ as value of the field at $x_0=0$ and $x_0=L_0+1$.
Note that viewed this way, the thickness of the film is $L_0+1$ rather 
than $L_0$. This provides a natural explanation of the result $L_s=1.02(7)$
obtained in \cite{myKTfilm} and might be a good starting point for a
field theoretic calculation of $L_s$.
The Hamiltonian of the two-component $\phi^4$ model, for a vanishing
external field, is given by
\begin{equation}
\label{hamiltonian}
{\cal H} = - \beta \sum_{<x,y>} \vec{\phi}_x \cdot \vec{\phi}_y
   + \sum_{x} \left[\vec{\phi}_x^2 + \lambda (\vec{\phi}_x^2 -1)^2   \right] \; ,
\end{equation}
where the field variable $\vec{\phi}_x$ is a vector with two real components. 
$<x,y>$ denotes a pair of nearest neighbour sites on the lattice.
The partition function is given by
\begin{equation}
Z =  \prod_x  \left[\int d \phi_x^{(1)} \,\int d \phi_x^{(2)} \right] \, \exp(-{\cal H}).
\end{equation}
Note that following the conventions of our previous work, e.g. \cite{ourXY}, 
we have absorbed the inverse temperature $\beta$ into the Hamiltonian.
In the limit $\lambda \rightarrow \infty$ the field variables are fixed to
unit length; i.e. the XY model is recovered. For $\lambda=0$ we get the exactly
solvable Gaussian model.  For $0< \lambda \le \infty$ the model undergoes 
a second order phase transition that belongs to the XY universality class.
Numerically, using Monte Carlo simulations and high-temperature series 
expansions, it has been shown that there is a value $\lambda^* > 0$, where 
leading corrections to scaling vanish.  Numerical estimates of $\lambda^*$
given in the literature are $\lambda^* = 2.10(6)$ \cite{HaTo99}, 
$\lambda^* = 2.07(5)$  \cite{ourXY} and most recently $\lambda^* = 2.15(5)$
\cite{recentXY}.  The inverse of the critical temperature $\beta_c$ has been 
determined accurately for several values of $\lambda$ using finite size
scaling (FSS) \cite{recentXY}. We shall perform our simulations at 
$\lambda =2.1$, since for this value of $\lambda$ comprehensive Monte 
Carlo studies of the three-dimensional system in the low and the 
high temperature phase have been performed 
\cite{myKTfilm,recentXY,myAPAM,myamplitude}.
At $\lambda =2.1$ one gets $\beta_c=0.5091503(6)$ \cite{recentXY}.
Since  $\lambda =2.1$  is not exactly equal to $\lambda^*$, there are 
still corrections $\propto L^{-\omega}$, although with a small amplitude.
In fact, following \cite{recentXY}, it should be by at least a factor
20 smaller than for the standard XY model.

\subsection{The energy and the specific heat}
\label{defineC}
First we should note that in eq.~(\ref{hamiltonian})  $\beta$ does not 
multiply the second term. Therefore, strictly speaking, $\beta$ is not 
the inverse temperature. However, 
in order to study universal quantities it is not crucial how the transition
line in the $\beta$-$\lambda$ plane is crossed, as long as this path is
smooth and not tangent to the transition line.
Here, following computational convenience, we vary $\beta$ at fixed $\lambda$.
Correspondingly we define the energy density as the derivative of the reduced
free energy density with respect to $\beta$. 
Furthermore we multiply by $-1$ to get positive numbers:
\begin{equation}
\label{Edef1}
E = \frac{1}{L_0 L_1 L_2} \frac{\partial \log Z}{\partial \beta}
 \end{equation}
It follows
\begin{equation}
\label{Edef}
 E =  \frac{1}{L_0 L_1 L_2}
 \left \langle  \sum_{<x,y>} \vec{\phi}_x \cdot \vec{\phi}_y \right \rangle \;.
 \end{equation}
We then define the specific heat as the derivative of the energy density with
respect to $\beta$:
\begin{equation}
\label{Cdef1}
C = \frac{\partial E}{\partial \beta} 
\end{equation}
It is easy to show that
\begin{equation}
\label{Cdef2}
 C = \frac{1}{L_0 L_1 L_2}
 \left (\left \langle \left [\sum_{<x,y>} \vec{\phi}_x \cdot \vec{\phi}_y
 \right ]^2 \right \rangle -
\left \langle \sum_{<x,y>} \vec{\phi}_x \cdot \vec{\phi}_y \right \rangle^2 
\right)  \;.
\end{equation}

\subsection{The correlation length}
The second moment correlation length $\xi_{2nd}$ and the transversal
correlation length $\xi_T$ of the bulk system are used to set the scale 
in the high and the low temperature phase, respectively.
Here we shall use the results  given in \cite{myamplitude}. For
completeness we recall the definitions of the two correlation lengths.

The second moment correlation length in k-direction is defined by
\begin{equation}
\xi_{2nd,k} \;=\; \sqrt{\frac{\chi/F_k-1}{4 \; \sin(\pi/L_1)^2}} \;\;\;,
\end{equation}
where
\begin{equation}
\label{chi}
 \chi =  \frac{1}{L_0 L_1 L_2} \langle \vec{M}^2 \rangle \;\;,
 \end{equation}
is the magnetic susceptibility and
\begin{equation}
F_k \;= \; \frac{1}{L_0 L_1 L_2} \;   \left \langle
\left |\sum_x \exp\left(i \frac{2 \pi x_k}{L_k} \right) \vec{\phi}_x \right |^2
\right \rangle
\end{equation}
is the Fourier transform of the correlation function at the lowest
non-zero momentum in k-direction.
Note that in the high temperature phase there is little difference between
$\xi_{2nd}$ and the exponential correlation length $\xi_{exp}$ which 
is defined by the asymptotic decay of the two-point correlation function.
Following  \cite{ourXY}:
\begin{equation}
\lim_{t\rightarrow 0} \frac{\xi_{exp}}{\xi_{2nd}} = 1.000204(3)  \;\;,\;\;\;\; (t<0)
\;\;
\end{equation}
for the thermodynamic limit of the three-dimensional system.
\footnote{Throughout, in the context of the 
$\phi^4$ model we use the convention $t=\beta-\beta_c$.}
In \cite{myKTfilm} we find for $\lambda=2.1$ by fitting the data for the 
second moment correlation length for $\beta \ge 0.49$:
\begin{equation}
\label{xires}
 \xi_{2nd} = 0.26362(8) (-t)^{0.6717}   \times [1 +0.039(8) (-t)^{0.527}
 - 0.72(4) (-t)] \;\;,
\end{equation}
where $t=\beta-0.5091503$.

The helicity modulus $\Upsilon$ gives the reaction of the system under
a torsion. To define the helicity modulus we consider a system, where
rotated boundary conditions are introduced in one direction (e.g. the 
1-direction):
For $x_1=L_1$ and $y_1=1$  the term $\vec{\phi}_x  \vec{\phi}_y$
in the Hamiltonian is replaced  by
\begin{equation}
\vec{\phi}_x \cdot R_{\alpha} \vec{\phi}_y =
\phi_x^{(1)} \left(\cos(\alpha) \phi_y^{(1)} + \sin(\alpha) \phi_y^{(2)} \right)
+\phi_x^{(2)}\left(-\sin(\alpha) \phi_y^{(1)} + \cos(\alpha) \phi_y^{(2)}\right) \;\;.
\end{equation}

The helicity modulus is then given by
\begin{equation}
\label{helidef}
\left . \Upsilon = - 
\frac{L_1}{L_0 L_2}
\frac{\partial^2 \log Z(\alpha)}{\partial \alpha^2} \right|_{\alpha=0} \;\;.
\end{equation}
 Note that we have skipped a factor of $T$ compared with the standard definition
 \cite{FiBaJa73}. Defined this way, the helicity modulus has the dimension of 
 an inverse length. 
In the literature $\xi_{\perp}=1/\Upsilon$ is referred to as 
transversal correlation length.
Fitting the data for the helicity modulus at $\lambda=2.1$,
given in table 2 of \cite{myamplitude}, up to $\beta=0.55$, we get
\begin{equation}
\label{Ups}
 \Upsilon = 1.5584(10) t^{0.6717}   \times (1 - 0.06(2) t)
 \end{equation}
with $t=\beta-0.5091503$.

\section{The finite size scaling behaviour of the specific heat}
\label{theory}
In this section we discuss the finite size scaling behaviour of 
the specific heat of thin films.  
The free energy density of the bulk system is given by
\begin{equation}
\label{bulkfree}
 f_{bulk}(t) = \tilde a t^{2 -\alpha} 
 \; (1+c_1 t^\theta + c_2 t^{2 \theta} + d_1 t^{\theta'} + e_1 t + \ldots) \; 
 + b(t)
\end{equation}
where $b(t)$ is the analytic background, $e_1 t$ is an analytic 
correction and $c_1 t^\theta$, $c_2 t^{2 \theta}$
and $d_1 t^{\theta'}$ are non-analytic corrections. 
In order to simplify the notation, we have omitted subscripts $\pm$ that 
indicate the phase.
Numerical values of
the corrections exponents are $\theta =\nu \omega \approx 0.527$ 
\cite{recentXY} and $\theta' \approx 1.2$ \cite{RG}.
Note that  the correction amplitudes $c_1$, $c_2$, $\ldots$ are small for
the $\phi^4$ model at $\lambda=2.1$, while $d_1$ and $e_1$ should assume 
generic values. In the following discussion we shall, for simplicity, 
ignore these corrections to scaling.
Inserting 
$ \xi(t)= \xi_0 t^{-\nu} $
into eq.~(\ref{bulkfree}) we arrive at
\begin{equation}
 f_{bulk}(t) = [\tilde a \xi_0^d]  \xi(t)^{-d} + b(t) 
\end{equation}
where we have used the hyperscaling relation $2-\alpha=d \nu$, where 
$d$ is the dimension of the system. Note that
$\tilde a \xi_0^d$ is universal. 

The free energy density of a thin film with periodic boundary conditions
with the thickness $L_0$ is given by \cite{Barber}
\begin{equation}
\label{periodicfree}
f_P(t,L_0) = L_0^{-d} q_P(t [L_0/\xi_0]^{1/\nu}) + b(t)
\end{equation}
where $d$ is the dimension of the system and
$b(t)$ in eq.~(\ref{periodicfree}) is the same function 
as in eq.~(\ref{bulkfree}). Also $q_P(x)$ is an analytic function at $x=0$.
There might be a singularity at  some $x \ne 0$ related with
the effectively two-dimensional transition.
In order to eliminate the analytic background one considers the difference 
\begin{eqnarray}
\label{xxxP}
f_{bulk}(t) - f_P(t,L_0) &=&
L_0^{-d} \left[ 
[\tilde a \xi_0^d]  (L_0/\xi(t))^{d} - q_P(t [L_0/\xi_0]^{1/\nu}) \right]
\nonumber \\
&=& L_0^{-d} p_P(t [L_0/\xi_0]^{1/\nu}) \;\;.
\end{eqnarray}
The specific heat is defined as minus the second derivative of $f$ with respect 
to an analytic function of $h(t)$. Using our definitions, it is minus the 
second 
derivative with respect to $t$ itself. Applied to eq.~(\ref{bulkfree}) we 
arrive at $A =- h'(0)^{-2}  (1-\alpha) (2-\alpha) \tilde a$ for the amplitude 
of the specific heat. Taking minus the second derivative of eq.~(\ref{xxxP})
with respect to $h(t)$ we arrive at
\begin{eqnarray}
\label{xxxC}
C_{bulk}(t) - C_P(t,L_0) &=& -h(0)^{-2}
 L_0^{-d} [L_0/\xi_0]^{2/\nu}  p_P''(t [L_0/\xi_0]^{1/\nu}) = \nonumber \\
- h(0)^{-2} \tilde a
[L_0/\xi_0]^{\alpha/\nu} [\tilde a \xi_0^d]^{-1}  p_P''(t [L_0/\xi_0]^{1/\nu})
&=& A^{-1} [L_0/\xi_0]^{\alpha/\nu} f_{2,P,R}(t [L_0/\xi_0]^{1/\nu}) \;\;.
\end{eqnarray}

In the case of free boundary conditions, which are studied in the present work, 
boundary effects have to be taken into account.
In \cite{myKTfilm} we have numerically shown that leading corrections can be
accounted for by replacing $L_0$ by $L_{0,eff} =L_0  + L_s$,
where we have obtained the estimate $L_s=1.02(7)$. Hence
\begin{equation}
\label{dirichletfree}
\frac{L_0}{L_{0,eff}} f_F(t,L_0) =
L_{0,eff}^{-d} q_F(t [L_{0,eff}/\xi_0]^{1/\nu})  
+ b(t) + c(t) L_{0,eff}^{-1} \;\;.
\end{equation}
The additional term $c(t) L_{0,eff}^{-1}$ gives a correction of the analytic 
background caused by the free boundary conditions.  Written 
this way it allows that
the correction to the background is given by an effective thickness that 
is different from $L_{0,eff}$. The prefactor $\frac{L_0}{L_{0,eff}}$ in front
of $f_F(t,L_0)$ corrects the volume that is used to compute the free energy
density.

Taking the same steps as in the case of periodic boundary conditions 
we arrive at
\begin{equation}
\label{definef2}
\left[C_{bulk}(t)-\frac{L_0}{L_{0,eff}}  C_F(t,L_0) \right]  
L_0^{-\alpha/\nu}= A \xi_0^{-\alpha/\nu}
f_{2,R}(t [L_{0,eff}/\xi_0]^{1/\nu}) -w(t) L_{0,eff}^{-1-\alpha/\nu}
\end{equation}

Alternatively in the literature one considers
\begin{equation}
\label{definef1}
\left[\frac{L_0}{L_{0,eff}}  C_F(t,L_0) - C_{bulk}(t_0)\right] L_0^{-\alpha/\nu}=
A \xi_0^{-\alpha/\nu} f_{1,R}(t [L_{0,eff}/\xi_0]^{1/\nu}) +w(t) L_{0,eff}^{-1-\alpha/\nu} 
\end{equation}
where  $\xi(t_0)=L_0$ or better $\xi(t_0)=L_{0,eff}$ for $t_0$ in the 
high temperature phase.  

Here we study the neighbourhood of the critical point. Therefore
we shall approximate $w(t) \approx w(0)$. To simplify the notation 
we shall write $w$ instead of $w(0)$ in the following.

Note that for large $|t| [L_{0,eff}/\xi_0]^{1/\nu}$ the difference  
$\left[C_{bulk}(t_0) - \frac{L_0}{L_{0,eff}}  C_F(t,L) \right]$
becomes small compared with $C_{bulk}(t_0)$ or $C_F(t,L)$.  On the 
other hand, corrections due to the boundary  are  
virtually independent on $|t| [L_{0,eff}/\xi_0]^{1/\nu}$. 
Therefore corrections due to the boundary might lead to  large 
relative errors for large values of $|t| [L_{0,eff}/\xi_0]^{1/\nu}$.

\section{Numerical Results}
\subsection{Thermodynamic limit of the three-dimensional system}
\label{threeD}
In this subsection we shall consider systems with periodic boundary 
conditions in all three directions and the linear size $L_0=L_1=L_2=L$. 
In the high temperature phase, corrections to the thermodynamic limit
decay exponentially with increasing lattice size. In practice
it turns out that for $L \gtrapprox 10 \xi_{2nd}$ these corrections 
are much smaller than the statistical error that we reach here.
Since a Goldstone mode  is present in the low temperature phase 
of the system \cite{HaLe1990,DiHaNaNi1991}  corrections to the 
thermodynamic limit decay only with some power of the linear lattice 
size. In particular for the energy density and the specific heat, 
we expect 
that, to leading order, the correction decays
$ \propto L^{-3}$.  
Therefore rather large lattices are needed to get a good approximation of
the thermodynamic limit.

One should note that the estimator~(\ref{Edef}) of the energy density is 
self-averaging, while the one~(\ref{Cdef2}) for the specific heat is not.
Since the lattices have to be rather large to avoid sizable finite 
size effects, it turns out that the specific heat can be most efficiently
determined by fitting the energy density in same range of $\beta$. 

To this end, we have computed the energy density for a large  number of 
$\beta$-values. As starting point we have taken the data given in 
tables 2 and 5 of \cite{myamplitude}.
These were supplemented by a rather large number of new simulations to
obtain a dense grid of $\beta$-values. In particular, we have  simulated
the  $96^3$ lattice in the range $0.521  \le \beta \le 0.58$ in steps of 
$\Delta \beta=0.0005$, for
$0.5025  \le \beta \le 0.5035$ in steps of $\Delta \beta=0.0001$ and
for $\beta=0.491$, $0.492$, $0.494$, $0.496$, $0.497$, $0.498$, $0.499$,
$0.501$, $0.502$, $0.503$ and $0.504$.
We simulated the
$128^3$ lattice at $\beta=0.516$, $0.517$, $0.518$, $0.519$ and $0.5205$, 
the 
$192^3$ lattice at $\beta=0.514$, $0.5145$, $0.5155$, $0.5165$, $0.5175$, 
$0.5185$ and  $0.5195$ and finally the 
$288$ lattice at $\beta=0.5115$, $0.5125$, $0.5135$, and $0.5145$. 

In the case of $L=96$ we typically performed $10^5$ measurements and 
$5 \times 10^4$ for the larger lattice sizes. For each of these measurements 
a Metropolis sweep, several overrelaxation sweeps and single cluster 
\cite{wolff}
updates were performed. For a discussion of the Monte Carlo algorithm
see \cite{myamplitude}.
In total, these simulations took a little less than one year of CPU time
on one core of a 2218 Opteron processor (2.60 GHz).

In \cite{myamplitude} the results for the thermodynamic limit in the 
low temperature phase were obtained by fitting the data of several lattice 
sizes with the ansatz $E(L) = E(\infty) + c L^{-3}$.  In the case of the 
simulations that we have added here, we have
checked that the $c L^{-3}$ corrections are sufficiently small to be ignored.

In the neighbourhood of the transition, we have fitted the energy density 
with the ansatz
\begin{equation}
\label{critical}
 E(\beta) = E_{ns} + C_{ns} (\beta-\beta_c) 
             + a_{\pm} |\beta-\beta_c|^{1-\alpha}
	     + d_{ns} (\beta-\beta_c)^2 
	     + b_{\pm} |\beta-\beta_c|^{2-\alpha}
\end{equation}
where $E_{ns}$, $C_{ns}$, $\beta_c=0.5091503(6)$ and 
$\alpha=-0.0151(3)$ \cite{recentXY} are input and
$a_{\pm}$, $d_{ns}$ and $b_{\pm}$ are the 5 free parameters of the fit. 

From the finite size scaling behaviour of $L^3$ systems with 
periodic boundary conditions in all directions  at the critical point
we find \cite{myAPAM}:
\begin{equation}
\label{e0l2.1}
 E_{ns} = 0.913213(5) + 20 \times (\beta_c- 0.5091503)
           + 5 \times 10^{-7} \times (1/\alpha+1/0.0151)
\end{equation}
for the non-singular part of the energy density and 
\begin{equation}
\label{c0l2.1}
 C_{ns} = 157.9(5) +  147000 \times (\beta_c-0.5091503)
 -2.1 \times (1/\alpha+1/0.0151)
\end{equation}
for the non-singular part of the specific heat at $\lambda=2.1$.

We did not include a term with the 
exponent $1-\alpha + \nu \theta' \approx 2$ into the ansatz~(\ref{critical}).
We expect that it is effectively taken into account by the last two terms
in~(\ref{critical}). Note that the main purpose of fitting  
the energy density with ansatz~(\ref{critical}) is to interpolate
our data in a large range of $\beta$-values.

After some preliminary studies, we decided to fit 
the energy density in the range $0.49 \le \beta \le  0.529$ using the 
ansatz~(\ref{critical}).
In total we have 98 data points in this interval and 
we get $\chi^2/$d.o.f. $=1.08$ for our fit. The results for the 
fit parameters are $a_+=160.688(2)$, $a_-=-151.459(2)$, 
$d_{ns}=-302.6(9.8)$, $b_+=302.4(10.3)$ and $b_-=293.4(10.3)$.

First  we have computed the universal combination 
\begin{equation}
R_{\alpha} = (1-A_+/A_-)/\alpha
\end{equation}
from the result of the fit. Note that $A_+/A_-=-a_+/a_-$. Using the 
central values for the input parameters we obtain $ R_{\alpha} = 4.035(16)$,
where we have taken into account the covariance of $a_+$ and $a_-$.
Furthermore we have checked the dependence of our result on the 
input parameters.
In fact, the dependence on the value of $\alpha$ is quite small. Taking 
the preferred value of the experiment on the space shuttle \cite{lipa2003}
$\alpha=-0.0127$ we get $R_{\alpha} = 4.022(16)$. We have also checked
the effect of the error of the other input parameters. It turns out that 
the uncertainty of $C_{ns}$ has the largest effect on $R_{\alpha}$:
Replacing $C_{ns}=157.9$  by $C_{ns}=158.4$  we get $R_{\alpha} = 4.025(16)$.
As our final result we quote
\begin{equation}
R_{\alpha} = 4.035(16)[20]
\end{equation}
where in $()$ we give the statistical error and in $[]$ the sum of all 
errors due to the uncertainty of the input parameters of the fit. Note
that the present result is compatible with the final result
$R_{\alpha} = 4.01(5)$ given in \cite{myAPAM}.

%
%
%
%
%

\begin{figure}
\begin{center}
\scalebox{0.64}
{
\includegraphics{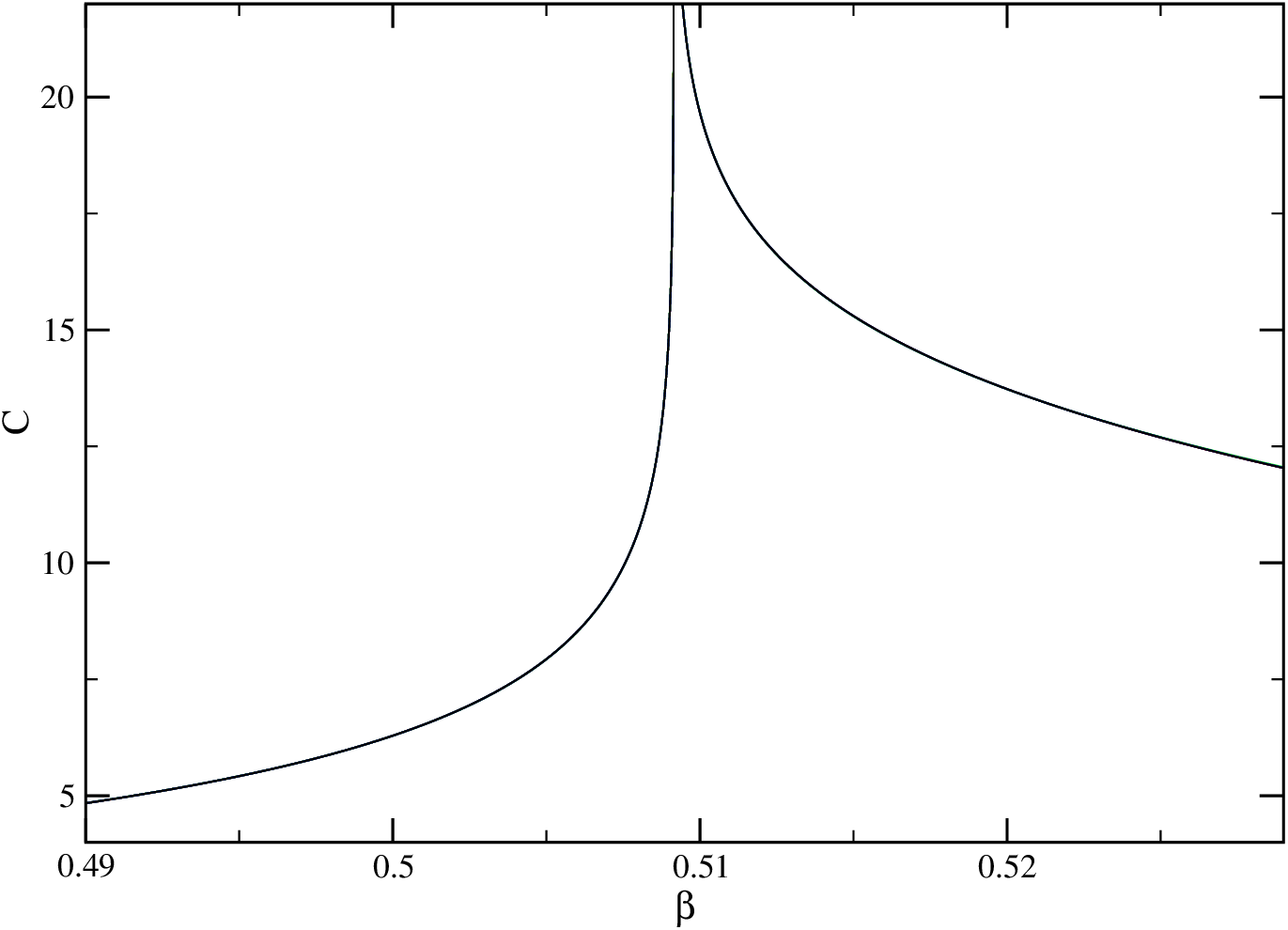}
}
\end{center}
\caption{
\label{C3D}  We plot our result for the specific heat
obtained from the fit of the energy density using ansatz~(\ref{critical}).
Note that at $\beta_c=0.5091503(6)$ the specific heat assumes the value
$C_{ns}=157.9(5)$, eq.~(\ref{c0l2.1}). In addition to the result obtained
by using the central values of the input parameters, we have also
plotted those were we have replaced the central value by the central value 
plus the error. For example $C_{ns}=157.9$ is replaced by $C_{ns}=158.4$.
At the resolution of the plot, all these curves fall on top of each other.
For a more detailed discussion see the text.
}
\end{figure}

\begin{figure}
\begin{center}
\scalebox{0.64}
{
\includegraphics{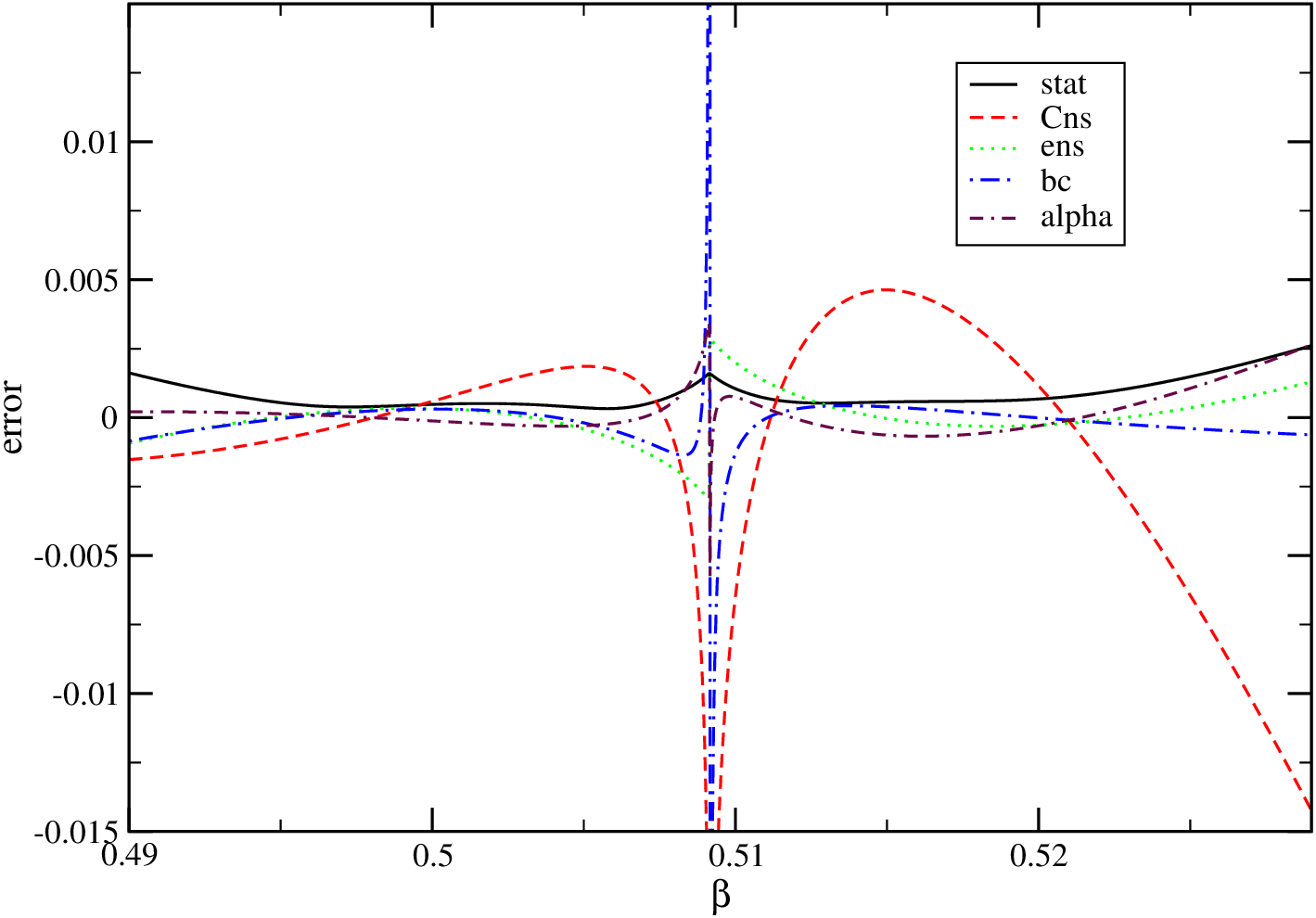}
}
\end{center}
\caption{
\label{ploterror}  Various sources of the error of the specific heat as 
obtained from fits of the energy density using ansatz~(\ref{critical}).
We plot the statistical error (stat) and the difference of the result
using the central values of the input parameters and results 
where we have replaced the central value of one of the input parameters 
by the central value plus the error.  In particular, we have replaced
$C_{ns}=157.9$ by $C_{ns}=158.4$  (Cns), $E_{ns}=0.913213$  
by  $E_{ns}=0.913218$
(ens), $\beta_c=0.5091503$  by $\beta_c=0.5091509$ (bc) and 
$\alpha=-0.0151$ by $\alpha=-0.0148$ (alpha). For a 
discussion see the text.
}
\end{figure}

In figure \ref{C3D} we have plotted the specific heat obtained from the 
fit of the energy density using ansatz~(\ref{critical}). Computing the 
statistical error, 
correlations among the fit-parameters are properly taken into account.
In this plot, errors can not be resolved.
%
%
%
%
%
In order to make the errors visible we have plotted in figure 
\ref{ploterror} the statistical error
and the difference between the results obtained  by using the central 
values of the input parameters and results where we have replaced 
one of the central values by the central value plus the error.

As one might expect, the differences diverge in the neighbourhood 
of the critical point.
In the low temperature
phase the largest uncertainty is due to the error of $C_{ns}$. In particular,
going to the upper boundary $\beta=0.529$ of our fit interval, 
the uncertainty induced by the error of $C_{ns}$ rapidly increases.
Therefore we decided to use the values for the specific heat 
obtained from the fit~(\ref{critical}) only
up to $\beta=0.525$. For larger values of $\beta$ we follow an alternative
approach as discussed below. In the case of the high temperature phase we shall
use the results obtained from the fit~(\ref{critical}) down to $\beta=0.49$.

In order to complete the computation of the specific heat, 
we have fitted our data for the energy density with the ansatz
\begin{equation}
\label{taylor}
 E(\beta) =  \sum_{i=0}^n a_i (\beta-\beta_0)^i 
\end{equation}
where $a_1$ is identified with the specific heat at $\beta_0$. 
We have included all values
of $\beta$ within the interval $[\beta_0- \Delta, \beta_0 + \Delta]$ into the 
fit. We have tested various values of $n$. Our final results are taken from 
fits with $n=4$.
In the range $0.525  < \beta_0 \le 0.529$ we have used $\Delta=0.008$.
For $0.529  < \beta_0 < 0.537$ we have used $\Delta=0.011$ 
and  for  $0.537  \le \beta_0 \le 0.58$  we have used $\Delta=0.016$.
In all cases $\chi^2/$d.o.f. is close to one.  To give an impression 
of the accuracy that is reached, we quote 
$C=12.690(3)$ at $\beta=0.525$ and  $C=11.899(2)$ at $\beta=0.53$.
  
For comparison with the results obtained from the fit with  
ansatz~(\ref{critical}) we have also performed fits with ansatz~(\ref{taylor})
in the range $0.522  \le \beta_0 \le 0.525$ using $\Delta=0.005$.
The numerical results for the specific heat obtained  from the two
different approaches are compatible within error bars. 
 
\subsection{Adjusting the scale of the axes}
\label{rescaleXY}
As we have mentioned already in the introduction, often in the literature 
the factor $A^{-1} \xi_0^{\alpha/\nu}$ is ignored when computing the 
finite size scaling functions $f_1$ and $f_2$.  Therefore, in order to
compare the results of different systems, we have to compute 
\begin{equation}
 r_{1,2} = \frac{[A \xi_0^{-\alpha/\nu}]_{system 1}}
                {[A \xi_0^{-\alpha/\nu}]_{system 2}} \;.
\end{equation}
In order to fix the ratio
between our definition for the specific heat of the $\phi^4$ model at 
$\lambda=2.1$ and that of
experiments on $^4$He at vapour pressure, we take the ratio of our result
for $A_-$ obtained in the previous subsection and that of   \cite{lipa2003}
for the three-dimensional thermodynamic limit of $^4$He at vapor pressure.
From table II of \cite{lipa2003} we read off  
\footnote{Note that our definition of $A_{\pm}$ and that used in  
\cite{lipa2003} differs by a factor of $\alpha$}
$\alpha A_- =  5.6537$.
Using an alternative fit ansatz the authors get $\alpha A_-  =  5.6950$.
We regard this difference as an estimate of the possible error of 
$\alpha A_- $. From the same fits, the authors obtain $\alpha=-0.01264$
and $-0.01321$, respectively. In order to match with these experimental 
numbers, we have taken $\alpha A_- =2.285$ obtained from 
fitting the energy density 
with  ansatz~(\ref{critical}), using $\alpha=-0.0127$ as input. 
Instead, using $\alpha=-0.0151$ we arrive at $\alpha A_- =2.322$. I.e. 
the value of $\alpha A_-$ is quite insensitive on the value of $\alpha$ that
is assumed.

Furthermore,
we need the amplitude of the correlation length for 
$^4$He at vapor pressure in the high and the low temperature phase. 
For the transversal correlation length in the low temperature phase
one finds \cite{SiAh84} and refs. therein:
\begin{equation}
\xi_T  \simeq 3.42 \AA \; t^{-\nu}
\end{equation}
where $t=1-T/T_{\lambda}$. Alternatively we can compute the 
amplitudes of the correlation lengths from $A_{\pm}$ using the results 
for the universal amplitude ratios
\begin{equation}
R_{\xi}^+=\xi_{0,2nd} (\alpha A_+)^{1/3} = 0.3562(10)  \;\;,
\end{equation}
\begin{equation}
R_{\xi}^-=\xi_{0,T} (\alpha A_-)^{1/3} = 0.850(5)
\end{equation}
and
\begin{equation}
R_{\Upsilon}=\frac{\xi_{0,2nd}}{\xi_{0,T}}=0.411(2)
\end{equation}
given in \cite{myamplitude}. To this end, we first have
to convert the results for the specific heat of the experiment from
$J \mbox{mole}^{-1} K^{-1}$ into $\AA^{-3}$.  To this end we need the  density 
$\rho_{\lambda} = 146.1087 \; \mbox{kg/m}^3$ \cite{KeTa64}
of $^4$He at the $\lambda$-transition and the Boltzmann  constant
$k_b=1.38065 \ldots \times 10^{-23} \mbox{J/K}$.
We arrive at 
\begin{equation}
\label{xi0T}
\xi_{0,T} = 3.45(3)  \AA
\end{equation}
where we have taken into account the errors of $R_{\xi}^-$ and of the 
experimental estimate of $\alpha A_-$.
In the case of the high temperature phase we get
\begin{equation}
\label{xi02nd}
\xi_{0,2nd} =  1.422(5) \AA
\end{equation}
where we have used 
\begin{equation}
\alpha A_+=1.05251 \times 5.6537 =  5.9506
\end{equation}
taken from table II of \cite{lipa2003} 
as input and the error is estimated by using 
\begin{equation}
\alpha A_+ = 1.05490 \times 5.6950 = 6.0077
\end{equation}
obtained from an alternative ansatz \cite{lipa2003}.

Now we are ready to compute the ratio
\begin{equation}
\left( \frac{\xi_{0,^4He}}{\xi_{0,\phi^4}} \right)^{-\alpha/\nu}
\end{equation}
where we can either use $\xi_{2nd}$ or $\xi_T$.  We get 
$1.0386$ using $\alpha=-0.0151$ or $1.0324$ using $\alpha=-0.0127$.

Hence we arrive at
\begin{equation}
\label{fuckor}
r_{^4He,\phi^4}  = 2.57
\end{equation}
with a relative uncertainty of about $2 \%$. Let us note that this number 
is only valid for $^4$He at vapour pressure and the $\phi^4$ model at 
$\lambda=2.1$ and the particular definitions of the specific heat that
have been used.

\subsection{Finite size scaling at $\beta_{c,3d}$}
\label{FSSTC}
First we performed simulations at the inverse transition
temperature $\beta_{c,3D}=0.5091503(6)$  \cite{recentXY}  
of the three-dimensional system. Here the correlation length of the 
thin film is relatively small; 
we find $\xi_{2nd,film}/L_{0,eff} \approx 0.416$. 
Hence already rather small ratios of $L_1/L_0$ are sufficient to approximate 
well the two-dimensional thermodynamic limit and therefore large values of 
$L_0$ can be 
reached. Furthermore, by construction $L_0/\xi_{3D}=0$.  Therefore this 
is an ideal location to accurately study the finite size scaling 
behaviour and in particular the corrections caused by  free boundary 
conditions. At the critical point of the three-dimensional system
eq.~(\ref{definef2}) reduces to
\begin{equation}
\label{CFbc}
\frac{L_0}{L_{0,eff}} C(t,L_0) = C_{ns} + c L_{0,eff}^{\alpha/\nu}
 + w L_{0,eff}^{-1} 
\end{equation}
where $C_{ns}=C_{bulk}(0)=157.9(5)$ \cite{myAPAM}  and 
$c=- f_2(0)$.

All numbers for the specific heat discussed in this section 
are determined  by using eq.~(\ref{Cdef2}).
As a first step, we have simulated the thickness $L_0=8$
for $L_1=L_2=16, 24, 32, 48$ and $64$. We conclude from these 
simulations that, at the level of our statistical error, 
the two-dimensional thermodynamic limit of the specific heat is reached for 
$L_1=L_2=48$. 
Based on this result we performed simulations
for $L_0=12,16,24,32,48$ and $64$  with $L_1=L_2=6 L_0$ throughout. 
In all cases,
we performed $10^6$ measurements, where for each measurement we performed one 
Metropolis sweep, two overrelaxation sweeps and wall cluster \cite{HaPiVi99}  
and single cluster updates.
The number of single cluster updates was chosen such that the average size of 
a cluster
times the number of clusters is a bit less than the number of lattice sites.
In total these simulations took about 5 month of CPU time on one core of 
a 2218 Opteron processor (2.60 GHz).
The results for the specific heat, as defined by eq.~(\ref{Cdef2}), 
are summarized in table \ref{heatbc}.

\begin{table}
\caption{\sl \label{heatbc}
Results for the specific heat $C$ obtained by using eq.~(\ref{Cdef2}) at  
$\beta=0.5091503$, which is the estimate of \cite{recentXY} for $\beta_{c,3D}$, 
 for lattices of the size $L_1=L_2=6 L_0$.
}
\begin{center}
\begin{tabular}{|r|r|}
\hline
\multicolumn{1}{|c}{$L_{0}$}&
\multicolumn{1}{|c|}{$C$} \\
\hline
 8 &    5.060(9)\phantom{0}  \\
12 &    6.145(13) \\
16 &    6.923(14) \\
24 &    8.062(17) \\
32 &    8.924(19) \\
48 &   10.141(24) \\
64 &   11.000(29) \\
\hline
\end{tabular}
\end{center}
\end{table}

We have fitted these data with ansatz~(\ref{CFbc}), where we have fixed
$C_{ns} =157.9$, $\beta_c=0.5091503$, $\alpha=-0.0151$  and $L_s=1.02$. The 
results of these fits are summarized in table \ref{fitheatbc}.
\begin{table}
\caption{\sl \label{fitheatbc}
Fitting the data for the specific heat given in table \ref{heatbc} using 
ansatz~(\ref{CFbc}).
Data for $L_{0,min} \le L_0 \le 64$  are included into the fit. We have fixed
$C_{ns}=157.9$, $\alpha=-0.0151$  and $L_s=1.02$.
}
\begin{center}
\begin{tabular}{|c|c|c|c|}
\hline
 $L_{0,min}$    &   $c$       &   $w$   & $\chi^2/$d.o.f. \\
\hline
\phantom{0}8  &--161.569(14)& 3.32(16)  &  1.01 \\
          12  &--161.597(20)& 3.90(33)  &  0.21  \\
          16  &--161.595(25)& 3.85(53)  &  0.28  \\
\hline
\end{tabular}
\end{center}
\end{table}
Already for $L_{0,min}=12$ the $\chi^2$/d.o.f. is smaller than one.
Going to larger $L_{0,min}$
the statistical error of $w$ rapidly increases.  Therefore we take the result
obtained for $L_{0,min}=12$ as our final result. Based on our data, 
it is impossible 
to give an estimate of systematic errors due to sub-leading corrections.

In order to check the dependence on the input parameters, 
we have repeated the fits
for $L_{0,min}=12$ using values of the input parameters that are 
shifted by the error of the input parameters:  E.g. in one of these 
fits $L_s=1.02$ is replaced by $L_s=0.95$, while the other input
parameters remain unchanged.
In the case of $\alpha$ and $\beta_c$, we have taken into account 
the effect of the shift on $C_{ns}$ and $E_{ns}$ as given by
eqs.~(\ref{e0l2.1},\ref{c0l2.1}).
It turns out that shifting $L_s$ has the largest effect on $w$.
Using $L_s=0.95$  we get $w=4.40(32)$. Taking the shifted value
$C_{ns}=158.4$ for the analytic background we get $w=4.17(33)$.  
Shifting the other input parameters has less impact on the value of 
$w$. 

Our result for $c=-f_2(0)$ can be compared with experiments and results
obtained by field theoretic methods.
On page 1028 in section V.A. of \cite{GaKiMoDi08} the authors 
analyse the scaling behaviour of the specific heat of thin films at the 
$\lambda$-transition of the three-dimensional system.   Fixing 
$\alpha=-0.01264$ they arrive at
\begin{equation}
 C(0,L_0) = [453.8 \pm 4.3] - [474.0 \pm 4.9] L_0^{-\alpha/\nu} 
\end{equation}
where $C$ is measured in units of $J \mbox{mole}^{-1} K^{-1}$ 
and $L_0$ in $\AA$.

Analysing our data for the specific heat at  $\beta_{c,3D}$, assuming 
$\alpha=-0.01264$, we arrive at $c  \approx - 190$.  Multiplying 
with $r_{^4He,\phi^4} = 2.57$  we arrive at $488$ in quite good agreement 
with the experimental result $[474.0 \pm 4.9]$, in particular when taking 
into account the error of $r_{^4He,\phi^4}$.

We can also write our result in terms of the universal ratio
\begin{equation}
-f_{2,R}(0)  = \frac{c}{A_+ \xi_{0,2nd}^{-\alpha/\nu} } = 1.0208(2)
-1.33 \times (\alpha+0.0151)
\end{equation}
where the error quoted in $()$ is dominated by the statistical error
of the specific heat of the thin films and the uncertainty of $L_s$. 
The dependence on the value of $\alpha$ that is used in the analysis is 
rather weak.
The authors of \cite{KrDi92} have computed 
$-f_{2,R}(0)$  (in their notation $\omega_{OO}$)
using  $\epsilon$-expansion to  $O(\epsilon)$.  Their result
is given in their eq.~(8.17).  To leading order 
$\omega_{OO}=1$, which is in quite good agreement with both 
our numerical result and with the experiment.  
Dohm and Sutter \cite{SuDo94} have pointed out  that the extrapolation 
to $\epsilon=1$ is affected by considerable ambiguities. In fact, setting 
$\epsilon=1$ the authors of \cite{KrDi92}  find $\omega_{OO} \approx 0.75$,
(table II of  \cite{KrDi92}) which is clearly ruled out by our result as
well as by experiment.

\subsection{Finite size scaling at $\beta_{KT}(L_0)$}
\label{FSSKT}
The KT phase transition occurs, up to scaling corrections, at a given 
value of the scaling variable $t L_0^{1/\nu}$ or equivalently $L_0/\xi_{T}$.
In \cite{myKTfilm} we find  $[L_0/\xi_{T}]^*=1.595(7)$. Following the KT theory
\cite{KT,Jo77,AmGoGr80}  the free energy is infinitely often 
differentiable with respect to the temperature at the KT transition.
Therefore no particular problem for the numerical analysis is expected. 
At the KT transition eq.~(\ref{definef2}) reduces to
\begin{equation}
\label{CKT}
C_{bulk}(\beta_{KT}(L_0))-\frac{L_0}{L_{0,eff}} C(\beta_{KT}(L_0),L_0)
=c_{KT} L_{0,eff}^{\alpha/\nu} - w L_{0,eff}^{-1}
\end{equation}
where $c_{KT}$ is $f_2$ at $[L_0/\xi_{T}]^*=1.595(7)$.

We have obtained accurate data for several values of $L_0$
in relation with  \cite{myKTfilm}. 
In table \ref{CbetaKT} we have summarized the results for the specific
heat obtained by using eq.~(\ref{Cdef2}) at the KT transition. 
 The results that are quoted were obtained
for $L_1=L_2=32 L_0$ lattices. In the case of $L_0=24$ and $32$, these are the 
largest available.  For $L_0<24$ we have checked that the 
results obtained from $L_1=L_2=32 L_0$ are consistent within the 
statistical error with those obtained from larger values of $L_1=L_2$.
In addition, we give numerical estimates for the specific heat of the  
three-dimensional bulk system as discussed in section \ref{threeD}.

\begin{table}
\caption{\sl \label{CbetaKT}
Results for the specific heat at the KT transition of thin films.
In the first column we give the thickness $L_0$ of the 
film, in the second column, we give the result for $\beta_{KT}(L_0)$ obtained 
in \cite{myKTfilm}.  In the third column we give the estimate of the 
specific heat $C_{bulk}$ of the three-dimensional bulk system at 
this temperature
and finally, in the fourth column the specific heat of the thin film at the 
KT transition. Throughout we have taken the result for 
$L_1=L_2=32 L_0$. For a detailed discussion see the text.
}
\begin{center}
\begin{tabular}{|r|l|l|l|}
\hline
\multicolumn{1}{|c}{$L_{0}$}&
\multicolumn{1}{|c}{$\beta_{KT}$} &
\multicolumn{1}{|c}{$C_{bulk}$} &
\multicolumn{1}{|c|}{$C_{film}$} \\
\hline
   6 &0.56825(1)[1]   & \phantom{0}8.586(1) &  \phantom{0}9.710(15) \\ 
   8 &0.549278(5)[9]  & \phantom{0}9.870(1) &            10.793(17) \\  
  12 &0.532082(3)[5]  &           11.617(2) &            12.179(20) \\
  16 &0.524450(2)[3]  &           12.791(7) &            13.176(23) \\ 
  24 &0.517730(2)[2]  &           14.342(5) &            14.454(22) \\  
  32 &0.514810(1)[2]  &           15.379(6) &            15.311(33) \\  
\hline
\end{tabular}
\end{center}
\end{table}

We have fitted these data with ansatz~(\ref{CKT}).
Our results, using the central values of the input parameters, 
are summarized  in table \ref{heatKT}.
\begin{table}
\caption{\sl \label{heatKT}
Fits of the specific heat $C$ at the KT transition 
with ansatz~(\ref{CKT}).
Data with $L_{0,min} \le L_0 \le 32$  are included into the fit. We have fixed
$\alpha=-0.0151$  and $L_s=1.02$.
}
\begin{center}
\begin{tabular}{|c|c|c|c|}
\hline
 $L_{0,min}$    &   $c_{KT}$       &   $w$   & $\chi^2/$d.o.f. \\
 \hline
\phantom{0}8  & 0.587(21) &  2.07(19)  & 2.98 \\
          12  & 0.638(27) &  2.81(31)  & 1.00 \\
          16  & 0.650(42) &  3.04(66)  & 1.42 \\
\hline
\end{tabular}
\end{center}
\end{table}
The value of $w$ is increasing with increasing minimal thickness $L_{0,min}$
that is included into the fit. The 
result obtained for $L_{0,min}=16$ is compatible within the statistical
error with that obtained in the preceding subsection for $\beta_{c,3D}$.

Next, we have checked how the results for $c_{KT}$ and $w$ depend on the 
values of the 
input for $L_s$ and $\alpha$. To this end we have repeated  fits for 
$L_{0,min}=16$ using shifted values for these input parameters.
Using $\alpha=-0.0151$ and $L_s=0.95$ as input we
get  $c_{KT}=0.634(42)$ and $w=3.61(65)$. Changing the value of the exponent
of the specific heat to $\alpha=-0.0148$ and keeping  $L_s=1.02$ we get
$c_{KT}=0.649(42)$ and $w=3.03(66)$. Finally we have set  
$\alpha=-0.0127$ obtained from the space shuttle experiment  \cite{lipa2003}.
Keeping $L_s=1.02$ we get  $c_{KT}=0.641(41)$ and $w=2.99(65)$.  

As one might expect, $w$ shows a strong dependence on the value of $L_s$. 
On the other hand the dependence of $w$ on $\alpha$ is quite weak;
even using the preferred value of \cite{lipa2003} the results for 
$c_{KT}$ and $w$  change only little. 
As our final estimate we take $c_{KT}=0.65(4)[2]$.  Taking into 
account both the analysis at $\beta_{c,3D}$ and at $\beta_{KT}$ we 
shall use $w=3.5$ in the following analysis of the specific heat in 
a large range of the scaling variable. The error of $w$ should be 
about $1$.

\subsection{Specific heat for $L_0=8$, $16$ and $32$ for a large range of $\beta$}
Finally we have computed the specific heat for $L_0=8$, $16$ and $32$ 
for a large range of $\beta$ in the neighbourhood of $\beta_{c,3D}$.
For this purpose, it turns out to be more efficient to compute the 
specific heat by taking the derivative of the energy density~(\ref{Edef}) 
with respect to $\beta$ numerically than by using eq.~(\ref{Cdef2}).

Let us first discuss how we have computed the two-dimensional 
thermodynamic limit of the energy density of the thin films.
In the high temperature phase of the thin film  we expect that the 
energy density converges exponentially fast with increasing $L=L_1=L_2$ 
toward the effectively two-dimensional thermodynamic limit.  
In order to see this asymptotic behaviour,
lattices with $L \gg \xi_{2nd,film}$ are needed.  From numerical results 
for the two-dimensional XY model we conclude that 
$L \gtrapprox  8 \xi_{2nd,film}$
is needed such that the deviation from the thermodynamic limit
is by far smaller than the statistical error that we typically reach in 
our study. 
For $L_0=8$, $16$ and $32$ we have simulated lattices up to $L=2048$,
$1800$ and $1024$, respectively. Therefore we could satisfy the condition 
$L \gtrapprox  8 \xi_{2nd,film}$ up to $\beta=0.545$, $\beta= 0.522$ and
$\beta= 0.5134$, respectively. At these values of $\beta$ we find 
$\xi_{2nd,film}=242.5(2)$, $153.16(15)$ and $111.99(15)$ for 
films of the thickness $L_0=8$, $16$ and $32$, respectively.





For $\beta > \beta_{KT}$ the asymptotic behaviour is given by the 
spin-wave approximation; i.e. by a free field theory. Therefore 
the thermodynamic limit is approached as $E(L) = E(\infty) + O(L^{-2})$. 
In this range of $\beta$ we have taken 
$E(\infty) = \frac{1}{3} [4 E(2 L) - E(L)]$ as our final result for 
the thermodynamic limit. 

Unfortunately there is a quite large range of $\beta$, where the 
extrapolation to the thermodynamic limit is less clear:
$\beta_{max} < \beta \le \beta_{KT}$, where  $\beta_{max}$ is the 
largest value of $\beta$ such that $L_{max} \gtrapprox  8 \xi_{2nd}$ and
$L_{max}$ is the largest lattice size that we can simulate with our 
(finite) computer resources.

In the context of \cite{myxi} and \cite{myrecentKT} we have simulated 
the two-dimensional XY model 
at $\beta=1.1199$, which is the best estimate of the inverse 
KT transition temperature \cite{HaPi97}, on lattices up to $L=4096$. 
Fitting the energy density for $16 \le L \le 4096$ with the ansatz
\begin{equation}
\label{effectiveansatz}
E(L) = E(\infty) + c L^{-\epsilon} \; 
\end{equation}
where $E(\infty)$, $c$ and 
$\epsilon$ are free parameters, we get $\epsilon = 1.85(2)$ with 
$\chi^2/$d.o.f.  smaller than one. 
Here we do not intend to further discuss this phenomenological observation;
I.e. whether this effective exponent is e.g. caused by logarithmic 
corrections.
We have also generated data for the 2D XY model for various values 
of $\beta$ in the range $\beta_{max} < \beta \le \beta_{KT}$ 
for various lattice sizes $L$  with $L \lessapprox 4 \xi_{2nd}$.   These data 
can be nicely fitted with the ansatz~(\ref{effectiveansatz}), where 
now $\epsilon$ apparently depends on $\beta$; it is decreasing with 
decreasing $\beta$. E.g.
for $\beta=1.0929$ we find, fitting the data for $16 \le L \le 512$, 
the effective exponent $\epsilon = 1.67(2)$. 
And for $\beta=1.0$, where $\xi_{2nd}=40.09(8)$, we find,
fitting the data for $16 \le L \le 128$, the effective 
exponent $\epsilon = 0.93(1)$.

From these observations in the two-dimensional XY model we learn that 
for  $\beta_{max} < \beta \le \beta_{KT}$ the extrapolation to the 
thermodynamic limit is non-trivial.   In order to keep  the systematic 
error small, we simulated as large lattices as possible. Since we had 
only few lattice sizes at hand, and the effective exponent $\epsilon$ is
not known a priori, we have also extrapolated our data in the range
$\beta_{max} < \beta \le \beta_{KT}$ using  $\epsilon=2$. In order
to estimate the systematic error of this extrapolation,
we compared results obtained for different $L$.

In table \ref{statistics} we have summarized the lattice sizes
$L_0$, $L=L_1=L_2$ and the values of $\beta$ 
that we have simulated. Typically we have performed $10^5$ measurements
for each simulation. Details are given in table \ref{statistics}.
For each measurement we performed one Metropolis sweep, several 
overrelaxation sweeps, single and wall-cluster updates. Integrated
autocorrelation times in units of measurements for the energy density are
$\tau_{E,int} \lessapprox 10$ for these simulations.
Note that in the case of $L_0=32$ we have skipped the range
$\beta_{max} \lessapprox \beta \lessapprox \beta_{KT}$,  since we  were
not able to simulate sufficiently large $L$ to get good control 
on the thermodynamic limit.
In total these simulations took about three years of CPU-time 
on a single core of a 2218 Opteron processor (2.60 GHz).

\begin{table}
\caption{\sl \label{statistics} We give a compilation of the simulations
that we have performed. In the first column we give the thickness $L_0$ of
the film. In the second column we give the size $L=L_1=L_2$ of the lattice 
in the other two directions. In the third and fourth column we give 
the upper and lower bound of the interval in the inverse temperature
that has been simulated. In the fifth column we give the step size 
$\Delta \beta$ that has been used. E.g. $\beta_{min}=0.49$, 
$\beta_{max}=0.52$ and $\Delta \beta=0.001$ means that 
$\beta=0.49, 0.491, 0.492$,
$\ldots$, $0.52$ have been simulated. Finally, in the last column we give 
the number of measurements (stat) that have been performed for each of the 
simulations.
}
\begin{center}
\begin{tabular}{|c|r|l|l|l|c|}
\hline
 \mc{1}{|c|}{$L_0$} &
\mc{1}{|c|}{$L_1=L_2$} & 
\mc{1}{|c|}{$\beta_{min}$}&
 \mc{1}{|c|}{$\beta_{max}$}& 
  \mc{1}{|c|}{$\Delta \beta$} &
   \mc{1}{|c|}{stat} \\
\hline
   8  &    64    & 0.49        & 0.52        & 0.001       & $5 \times 10^5$ \\
   8  &   128    & 0.52        & 0.527       & 0.001       & $2 \times 10^5$ \\
   8  &   128    & 0.544       & 0.56        & 0.001       & $2 \times 10^5$ \\
   8  &   256    & 0.528       & 0.537       & 0.001       & $10^5$ \\
   8  &   256    & 0.544       & 0.56        & 0.001       & $10^5$ \\
   8  &   256    & 0.562       & 0.58        & 0.002       & $10^5$ \\
   8  &   512    & 0.5375      & 0.548      & 0.0005      &  $10^5$ \\
   8  &   512    & 0.549       & 0.56       & 0.001       &  $10^5$ \\
   8  &   512    & 0.562       & 0.58       & 0.002       &  $10^5$ \\
   8  &  1024    & 0.5435      & 0.548      & 0.0005      &  $10^5$ \\
   8  &  2048    & 0.545       &            &             &  $14 \times 10^4$ \\
\hline
  16  &   128    & 0.49        & 0.4995      & 0.0005      & $10^5$ \\
  16  &   256    & 0.527       & 0.550       & 0.001       & $2 \times 10^5$ \\
  16  &   512    & 0.500       & 0.512       & 0.001       & $10^5$ \\ 
  16  &   512    & 0.5125      & 0.529       & 0.0005      & $10^5$ \\
  16  &   512    & 0.53       &  0.55       & 0.001       & $10^5$ \\
  16  &  1024    & 0.521       & 0.529       & 0.0005      & $10^5$ \\
  16  &  1600    & 0.5223      &             &             & $10^5$ \\
  16  &  1800    & 0.522       & 0.523       & 0.00025     & $5 \times 10^4$ \\
\hline
  32  &   256    & 0.49        & 0.505       & 0.0005      & $10^5$ \\
  32  &   256    & 0.50525     & 0.50875     & 0.00025     & $10^5$ \\
  32  &   256    & 0.516       & 0.518       & 0.0002      & $10^5$ \\
  32  &   256    & 0.5185      & 0.526       & 0.0005      & $10^5$ \\
  32  &   512    & 0.509       & 0.5128      & 0.0002      & $2 \times  10^5$\\
  32  &   512    & 0.516       & 0.518      & 0.0002      & $ 10^5$ \\
  32  &   512    & 0.5185      & 0.526      & 0.0005      & $ 10^5$ \\
  32  &  1024    & 0.513       & 0.5136     & 0.0002      & $5 \times  10^4$\\
\hline
\end{tabular}
\end{center}
\end{table}

Similar to the case of the three-dimensional bulk system in the low 
temperature phase, we have computed the specific heat by fitting the 
data of the energy density with the ansatz~(\ref{taylor}). Also here
we have taken our final results from fits with $n=4$. We have adjusted the 
range of the fit $\Delta$ such that $\chi^2/$d.o.f. is about one. 

In the case of $L_0=8$ we have used 
in the neighbourhood of the maximum of the specific heat 
$\Delta =0.003$.
As we go away from the maximum $\Delta$ is increased up to 
$\Delta=0.011$ for the smallest and largest values of $\beta$ 
that we have simulated.

\begin{figure}
\begin{center}
\scalebox{0.52}
{
\includegraphics{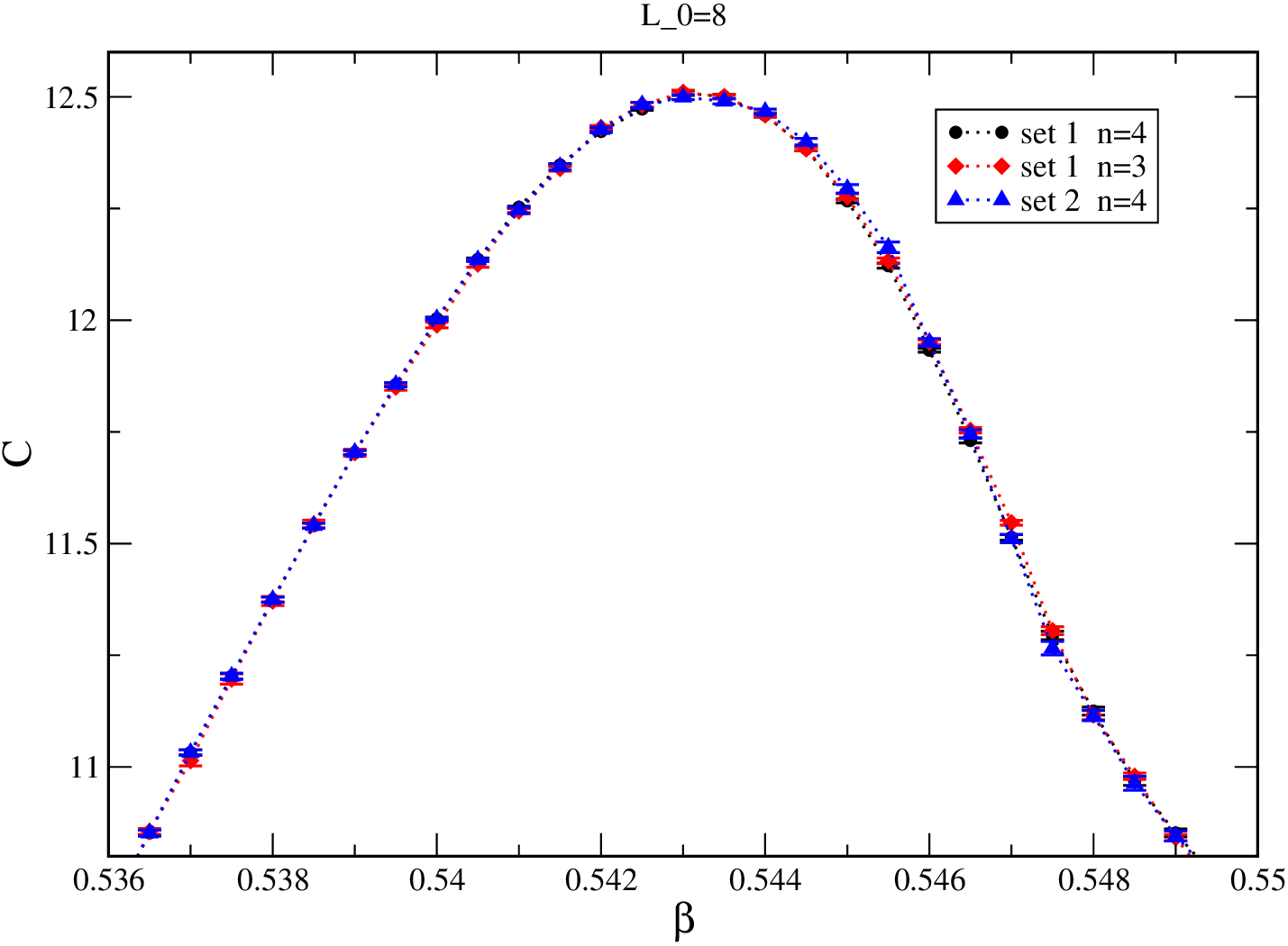}
}
\end{center}
\caption{
\label{C8vgl}  In the figure we give results for the thermodynamic 
limit of the specific heat  defined by eq.~(\ref{Cdef1})
for films of the thickness $L_0=8$.  These  results were obtained by
fitting our data for the energy density with the ansatz~(\ref{taylor}) using 
$n=3$ and $n=4$. We have used two different sets (set 1, set 2) of data for 
the energy density.
For a detailed discussion see the text.  Note that 
$\beta_{KT}=0.549278(5)[9]$ for $L_0=8$, while the maximum of the specific heat
is located at $\beta\approx 0.5432$ as we read off from the plot.
}
\end{figure}

In figure \ref{C8vgl} we plot results for the specific heat for $L_0=8$
in the most 
difficult range of $\beta$ around the maximum of the specific heat.
With this plot we like to check two sources of systematic error: 
First the truncation effects of eq.~(\ref{taylor}). To this end we have 
plotted the results obtained with $n=3$ in addition to those obtained with 
$n=4$. Second we investigate systematic errors of the extrapolation 
to the two-dimensional thermodynamic limit in the range 
$\beta_{max} < \beta \le \beta_{KT}$. To this end we have replaced the 
values for the energy density obtained from the extrapolation of $L=512$ and
$L=1024$  (set 1) by those obtained from the extrapolation of 
$L=256$ and $L=512$ (set 2).

The results obtained from $n=3$ and $n=4$ fall nicely on top of each other.
Hence there should be no systematic errors due to the truncation of 
eq.~(\ref{taylor}) that are considerably larger than the statistical error. 
On the 
other hand, comparing the results  obtained from set 1  and set 2 we see
discrepancies that are a few times the statistical error. Note that this 
problem affects only  the range  $\beta_{max} < \beta \le \beta_{KT}$.
The statistical error of the specific heat is maximal at the peak of the 
specific heat. There it is about $0.01$.
Instead, computing the specific heat by using eq.~(\ref{Cdef2})
 we get e.g.
for $\beta=0.543$, $L=512$ the result $C=12.52(9)$, 
for $\beta=0.5435$, $L=512$ the result $C=12.62(9)$ and 
for $\beta=0.5435$, $L=1024$  the result $C=12.48(11)$.  I.e. the 
values for the specific heat are compatible with those obtained by fitting 
the energy density, however the statistical error is about ten times larger.

In the case of $L_0=16$ we have used in the neighbourhood of the peak of the 
specific heat $\Delta=0.001$ for the fits of the energy density. 
As we go away from the maximum $\Delta$ is increased up to
$\Delta=0.015$ for the smallest and largest values of $\beta$
that we have simulated. At the maximum of the specific heat the statistical
error is about $0.01$. The ratio $L/L_0$ that we have maximally reached 
for $L_0=16$ is smaller than  for $L_0=8$. It is comparable with that of 
the data included into set 2 for $L_0=8$ discussed above. Therefore we 
expect that deviations from the two-dimensional thermodynamic limit in the 
range $\beta_{max} < \beta \lessapprox \beta_{KT}$ are of similar size 
as for set 2, i.e. a few times the statistical error that we have reached.
Note again that outside of this interval, the two-dimensional 
thermodynamic limit is well under control.

In the case of $L_0=32$ we  did not study the range 
$\beta_{max} < \beta \lessapprox \beta_{KT}$, since we could not simulate
sufficiently large lattices to get a good approximation of the 
two-dimensional thermodynamic limit.  For the $\beta$-values closest to 
the peak of the specific heat, we have used $\Delta=0.0015$. For our 
largest and smallest values of $\beta$ we have used $\Delta=0.015$. 
Close to the peak the statistical error of $C$ is about $0.01$.

\begin{figure}
\begin{center}
\scalebox{0.62}
{
\includegraphics{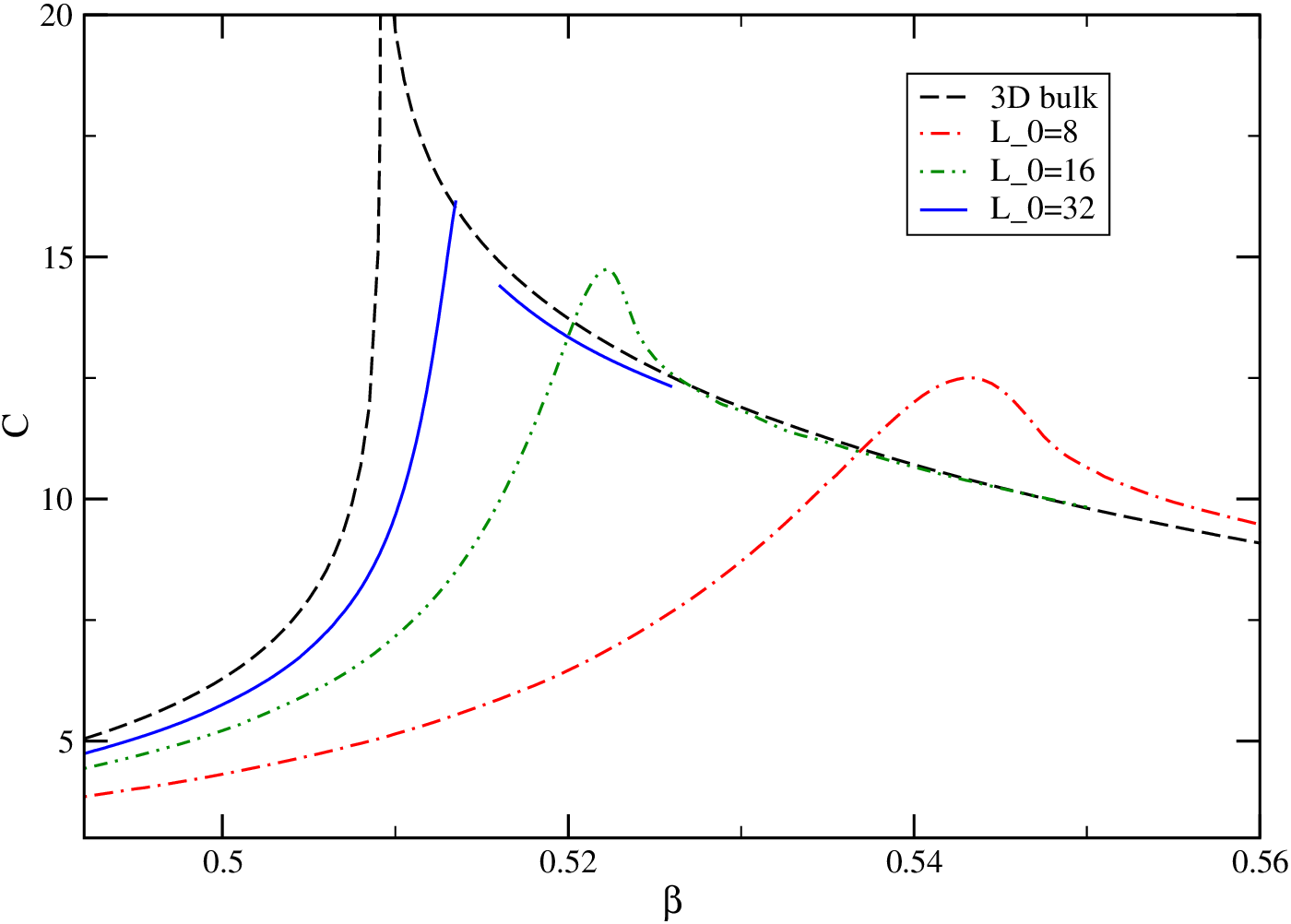}
}
\end{center}
\caption{
\label{rawC}  We plot our results for the
specific heat as a function of the inverse temperature $\beta$.
In addition to the 3D bulk specific heat we give our results for the 
thicknesses $L_0=8$, $16$ and $32$.
}
\end{figure}

In figure \ref{rawC} we have plotted our results for the specific heat
for the three-dimensional bulk system and the thicknesses $L_0=8$, $16$ and 
$32$. 

The position of the 
peak of the specific heat approaches the transition temperature of 
the three-dimensional bulk system and the hight of the peak increases as
the thickness $L_0$ of the film increases.  It is interesting to note 
that in the case of $L_0=8$ for low temperatures the specific heat 
of the film is larger than for the bulk system, while for $L_0=32$  it is 
smaller. The specific heat of the films at the maximum is larger than 
the bulk specific heat.  

\section{Results for the finite size scaling function $f_2$}
In this section we compute the finite size scaling functions $f_2$  using
the numerical data for the specific heat discussed above.  Our
results are compared with those obtained from experiments on films of 
$^4$He near the $\lambda$-transition and with results obtained by using 
field theoretic methods.
\subsection{The high temperature phase}
\begin{figure}
\begin{center}
\scalebox{0.62}
{
\includegraphics{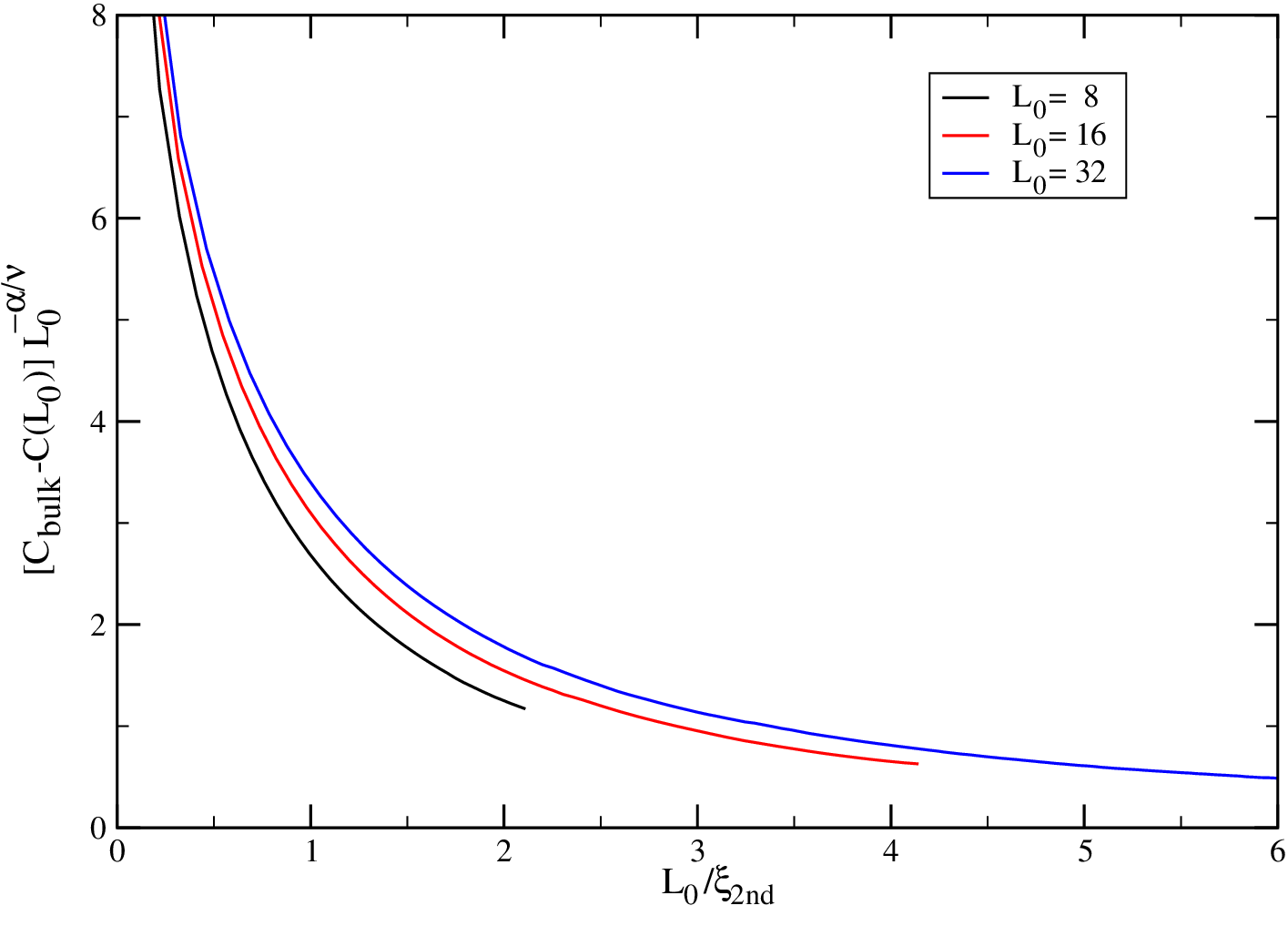}
}
\end{center}
\caption{
\label{F2NH}  We plot
$[C_{bulk}(\beta)-C(\beta,L_0)] L_{0}^{-\alpha/\nu}$
as a function of $L_0/\xi_{2nd}(\beta)$
for the thicknesses $L_0=8, 16$ and $32$.
For a discussion see the text.
}
\end{figure}

First we compute $f_2$ without taking into account
corrections. 
To this end, in figure \ref{F2NH}, we have plotted 
$[C_{bulk}(\beta)-C(\beta,L_0)] L_{0}^{-\alpha/\nu}$ as a function 
of $L_{0}/\xi_{2nd}(\beta)$, where $\xi_{2nd}(\beta)$ is given by 
eq.~(\ref{xires}). Corrections are clearly visible.
Note that in the whole range of $L_{0}/\xi_{2nd}(\beta)$ that is plotted,
the error of $[C_{bulk}(\beta)-C(\beta,L_0)]$ should be at most $0.01$.
I.e. the error is much smaller than the difference between the results
obtained for different $L_0$.

\begin{figure}
\begin{center}
\scalebox{0.62}
{
\includegraphics{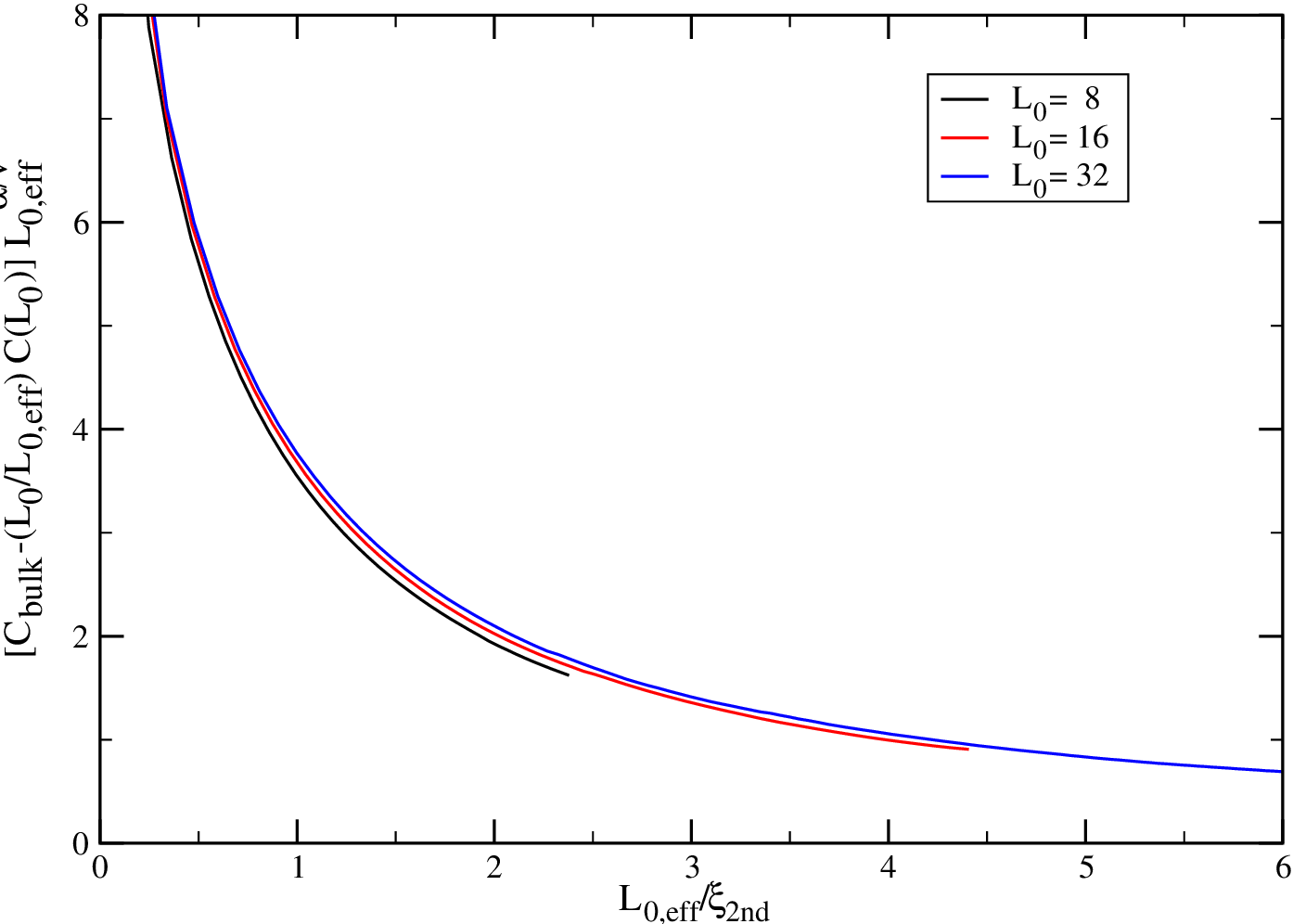}
}
\end{center}
\caption{
\label{F2EH}  We plot
$[C_{bulk}(\beta)-C(\beta,L_0) L_0/L_{0,eff}] L_{0,eff}^{-\alpha/\nu}$
as a function of $L_{0,eff}/\xi_{2nd}(\beta)$
for the thicknesses $L_0=8, 16$ and $32$.
For a discussion see the text.
}
\end{figure}

Next, in figure \ref{F2EH}, we have taken into account 
boundary corrections by replacing $L_0$ by $L_{0,eff}=L_0+L_s$ with 
$L_s=1.02$.  
This has two effects:  On the $x$-axis we
replace $L_0/\xi_{2nd}$ by $L_{0,eff}/\xi_{2nd}$ and secondly, in order to
compute the energy density, we replace the volume $L_0 L_1 L_2$ by
the effective volume $L_{0,eff} L_1 L_2$. Hence the specific heat that
we have computed before is multiplied by $L_0/L_{0,eff}$.
Now the corrections are much reduced in comparison with figure \ref{F2NH}.
In particular the curves for $L_0=16$ and $L_0=32$ fall almost on top of 
each other.

\begin{figure}
\begin{center}
\scalebox{0.62}
{
\includegraphics{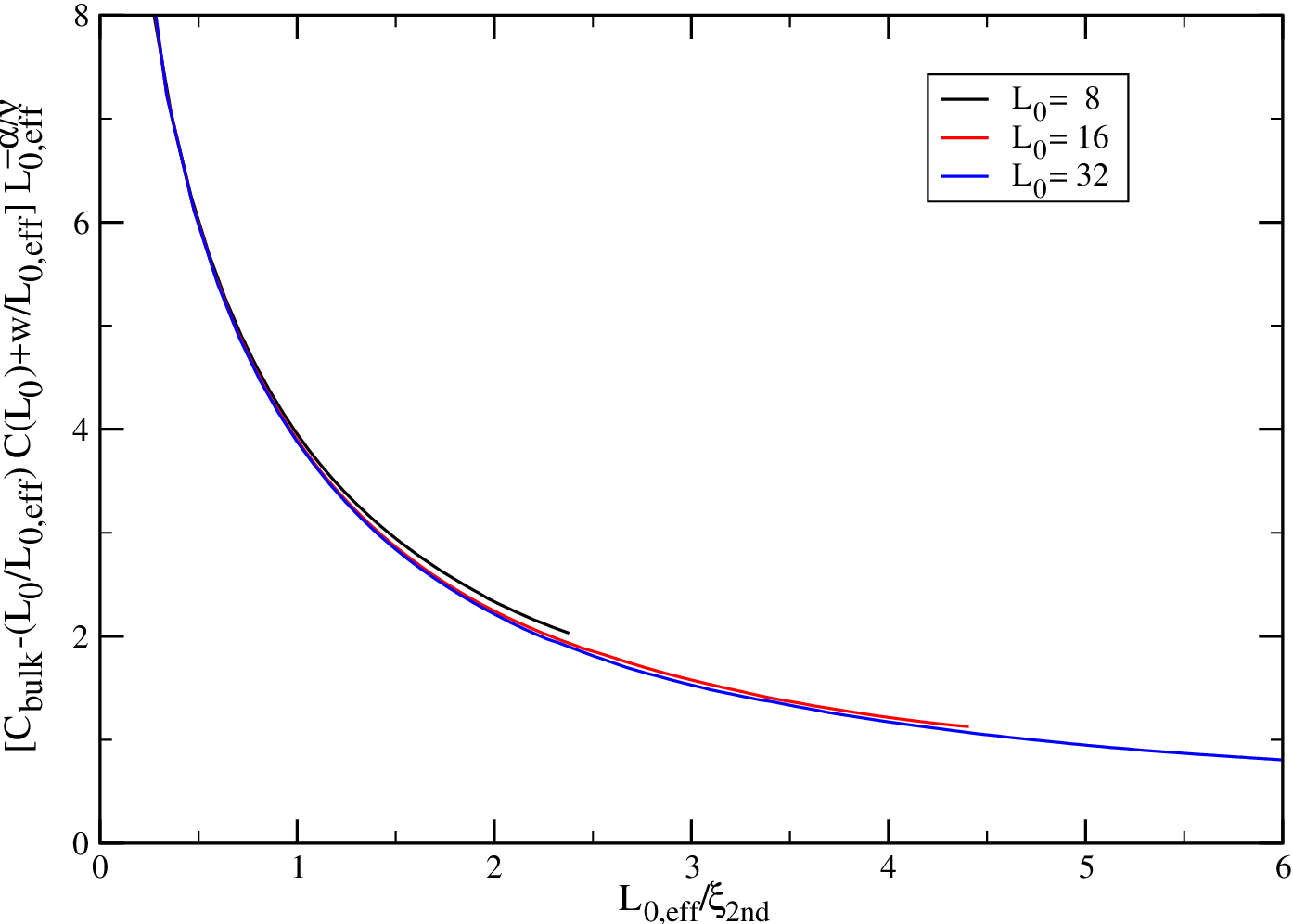}
}
\end{center}
\caption{
\label{F2EWH}  In the figure we plot
$[C_{bulk}(\beta)-C(\beta,L_0) L_0/L_{0,eff} +w/L_{0,eff}] L_{0,eff}^{-\alpha/\nu}$
with $w=3.5$ as a function of $L_{0,eff}/\xi_{2nd}(\beta)$
for the thicknesses $L_0=8, 16$ and $32$.
For a discussion see the text.
}
\end{figure}

Finally, in figure \ref{F2EWH},
we have taken into account corrections due to possible 
boundary effects on the analytic background by adding 
the term $w/L_{0,eff}$. 
The curves match only marginally better than in figure \ref{F2EH}. 
Note that while in the case of figure \ref{F2EH} the value of 
$f_2$ was 
increasing with increasing $L_0$, it is decreasing in the case of 
figure \ref{F2EWH}. Therefore it seems to be reasonable to assume 
that the asymptotic result for $f_2$ is located between 
$[C_{bulk}(\beta)-C(\beta,L_0) L_0/L_{0,eff}] L_{0,eff}^{-\alpha/\nu}$  and 
$[C_{bulk}(\beta)-C(\beta,L_0) L_0/L_{0,eff}
+w/L_{0,eff}] L_{0,eff}^{-\alpha/\nu}$ with $w=3.5$.

In table \ref{f2numbers} we give $f_2$ for a few
values of $L_{0,eff}/\xi_{2nd}$. This should help the reader to compare
our result with that obtained from other systems.

\begin{figure}
\begin{center}
\scalebox{0.62}
{
\includegraphics{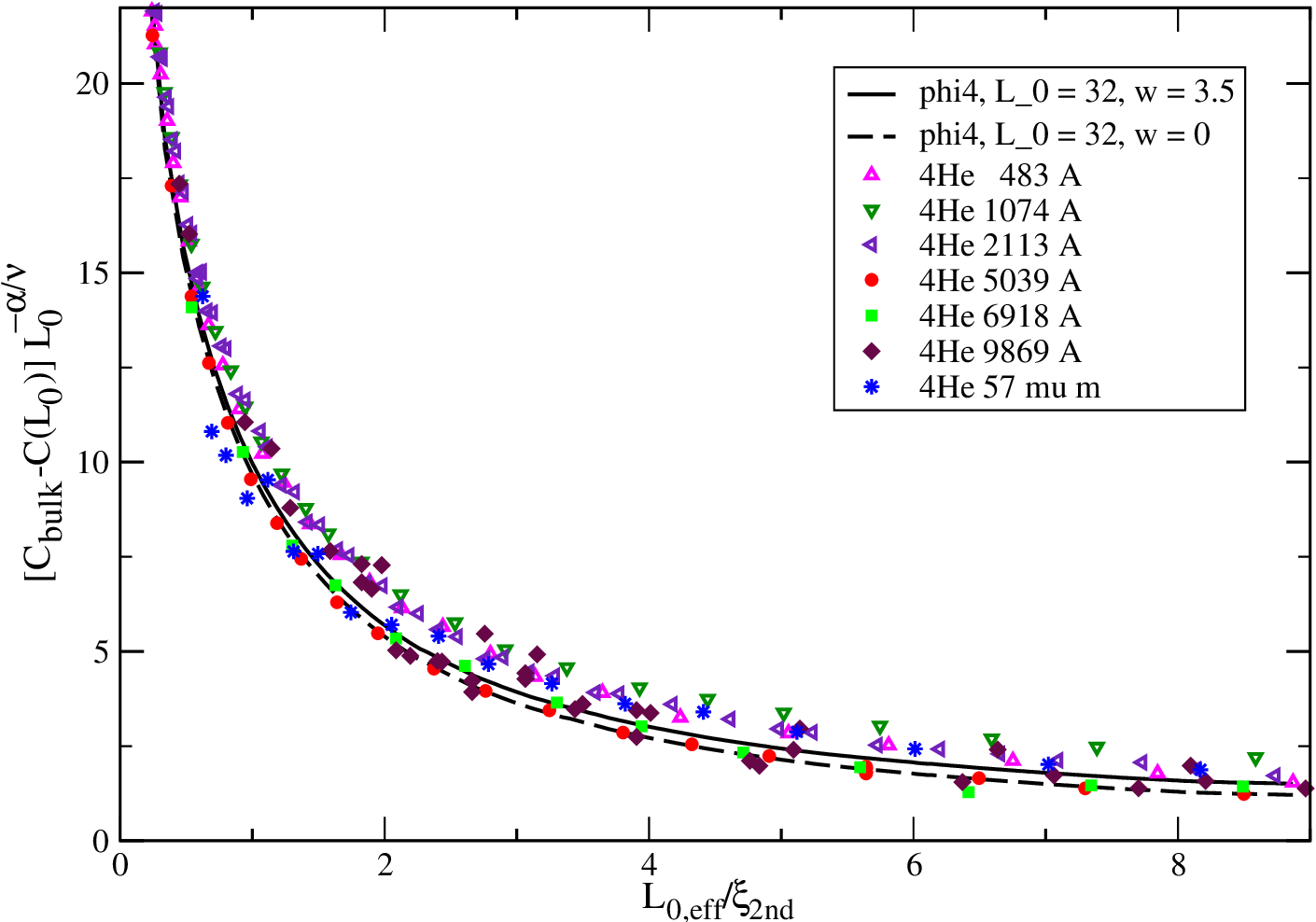}
}
\end{center}
\caption{
\label{F2EXPH} Experimental results 
\cite{KiMeGa99,KiMeGa00,Lietal00} obtained for films
of $^4$He in the high temperature phase, close to the $\lambda$-transition.  
We plot $[C_{bulk}(t)-C(t,L_0)] L_{0}^{-\alpha/\nu}$ as a function of
$L_0/\xi_{2nd}(t)$ for the thicknesses $L_0=483$,
$1074$, $2113$, $5039$, $6918$, $9869 \AA$ and $57$ $\mu$m. 
For comparison we give our results obtained for $L_0=32$. In contrast to 
the experimental results, boundary corrections, characterized by $L_s=1.02$ 
and $w=3.5$ (solid black line) or by $L_s=1.02$ and $w=0$ (dashed black line) 
are taken into account. Furthermore we have multiplied our results for 
$f_2$ by $r_{^4He,\phi^4}=2.57$. 
For a discussion see the text.
}
\end{figure}

\begin{table}
\caption{\sl \label{f2numbers}
We give our
results for $f_2$ obtained from $L_0=32$ and 
corrections characterized by $L_s=1.02$ and $w=3.5$ for a few
values of $L_{0,eff}/\xi_{2nd}$ (upper row).
Using instead $L_s=1.02$ and $w=0$, the value
of $f_2$ is smaller by about $0.11$ throughout. 
}
\begin{center}
\begin{tabular}{|ccccccccccccc|}
\hline
0.8 & 0.9 & 1.0  & 1.1  & 1.2 & 1.3 & 1.4 & 1.5 & 1.6 & 1.7 & 1.8 & 1.9 & 2.0\\
4.52 & 4.17 & 3.88 & 3.62 & 3.39 & 3.19 & 3.00 & 2.84 & 2.69 & 2.56 & 2.43 & 
 2.32 & 2.21 \\
\hline
\end{tabular}
\end{center}
\end{table}
In figure \ref{F2EXPH} we plot  $[C_{bulk}(t)-C(t,L_0)] L_0^{-\alpha/\nu}$
as a function of $L_0/\xi_{2nd}(t)$ using the experimental data of  
\cite{KiMeGa99,KiMeGa00}  for thin films of $^4$He at vapor pressure 
of the thicknesses
483, 1074, 2113, 5039, 6918  and $9869 \AA$. These data 
are taken from the web page \cite{Gaspariniweb}.
The experimental data for the specific heat are given as a function of the 
reduced temperature $t=(T-T_{\lambda})/T_{\lambda}$. In order to plot
them as a function of $L_0/\xi_{2nd}(t)$ we use
$\xi_{2nd}=1.422 \; |t|^{-\nu} \AA$, eq.~(\ref{xi02nd}). As value of the 
critical exponents we take
$\nu=0.6717$ \cite{recentXY} and correspondingly $\alpha=-0.0151$.
Furthermore we plot the results of \cite{Lietal00} for a film of 57 $\mu$m
as thickness.
The numbers \cite{LipaPrivate} used for the plot are those  of
\cite{BaHaLiDu07} plotted in figure 29. 
We have replaced $x=t L_0^{1/\nu}$ by $L_0/\xi_{2nd}$  on the x-axis
and have multiplied $f_2$ by a factor 
$57000^{0.0151/0.6717-0.0127/0.6709}=1.04$. Note that the authors of 
\cite{BaHaLiDu07} assume $\nu=0.6709$ \cite{lipa2003}, while we
prefer $\nu=0.6717$  \cite{recentXY}.

For comparison we give
our result for $f_2$ obtained from $L_0=32$, 
taking into account boundary
corrections characterized by $L_s=1.02$ and $w=3.5$ (solid black line). 
To indicate the 
possible error we give in addition the result
obtained for $L_0=32$, $L_s=1.02$ and $w=0$ (dashed black line).
We have multiplied our numbers by $r_{^4He,\phi^4}=2.57$, eq.~(\ref{fuckor}).

We observe that $f_2$ computed from the specific heat of 
$483, 1074$ and $2113$ $\AA$ films
is systematically larger than our result.  In contrast, we see a quite good 
match  with the results obtained from $5039$ and $6918$ $\AA$  and (a 
little worse) for $9869 \AA$ films.  Also in the case of the $57$ $\mu$m film 
we see a reasonable match with our result.
Note that there are experimental results available for much larger
and for smaller values of $L_0/\xi$ than plotted in figure \ref{F2EXPH}.

The scaling function $f_2$ has been calculated perturbatively to one-loop 
in three dimensions fixed \cite{ScWaDoFr90,Dohm93}. Also in this case,
the relative factor between the experimental and theoretical results for 
the specific heat was fixed by using the behaviour of the specific heat in the 
thermodynamic limit. To this end the experimental data of \cite{TaAh85} had 
been used.  In \cite{ScWaDoFr90} the result for $f_2$ is only given 
as log-log plot. Therefore we abstain from plotting it  in
figure \ref{F2EXPH}. E.g. in figures 29 and 31 of \cite{BaHaLiDu07} 
the field theoretic result of \cite{ScWaDoFr90} is plotted along with 
the experimental results of \cite{lipa2003}.  There is a quite reasonable
match between the field theoretic result and the experimental data.
For $t L_0^{1/\nu} \lessapprox 12$, the experimental result for $f_2$ is 
somewhat larger than that of \cite{ScWaDoFr90}. To pick out
one point:
For $t L_0^{1/\nu} = 1$ we read off from figure 31 of  \cite{BaHaLiDu07}
for \cite{ScWaDoFr90} the value $f_2 \approx 10$. This has to be compared 
with our result $f_2(\xi_{2nd}/L_0=1/1.422) = 12.6(1)$, where the factor 
$r_{^4He,\phi^4}=2.57$ has been taken into account.

In eq.~(8.1) of
 \cite{KrDi92} the specific heat of the thin film has been calculated
up to O$(\epsilon)$ as a function of $L_0/\xi_{2nd}$. Here we abstain from 
evaluating this function. In section \ref{FSSTC} we have discussed 
the case $L_0/\xi_{2nd} = 0$. Below, in section 
\ref{larget}, we shall discuss the limits $L_0/\xi_{2nd} \rightarrow \infty$
and $L_0/\xi_{T} \rightarrow \infty$.

\subsection{The low temperature phase}
\begin{figure}
\begin{center}
\scalebox{0.62}
{
\includegraphics{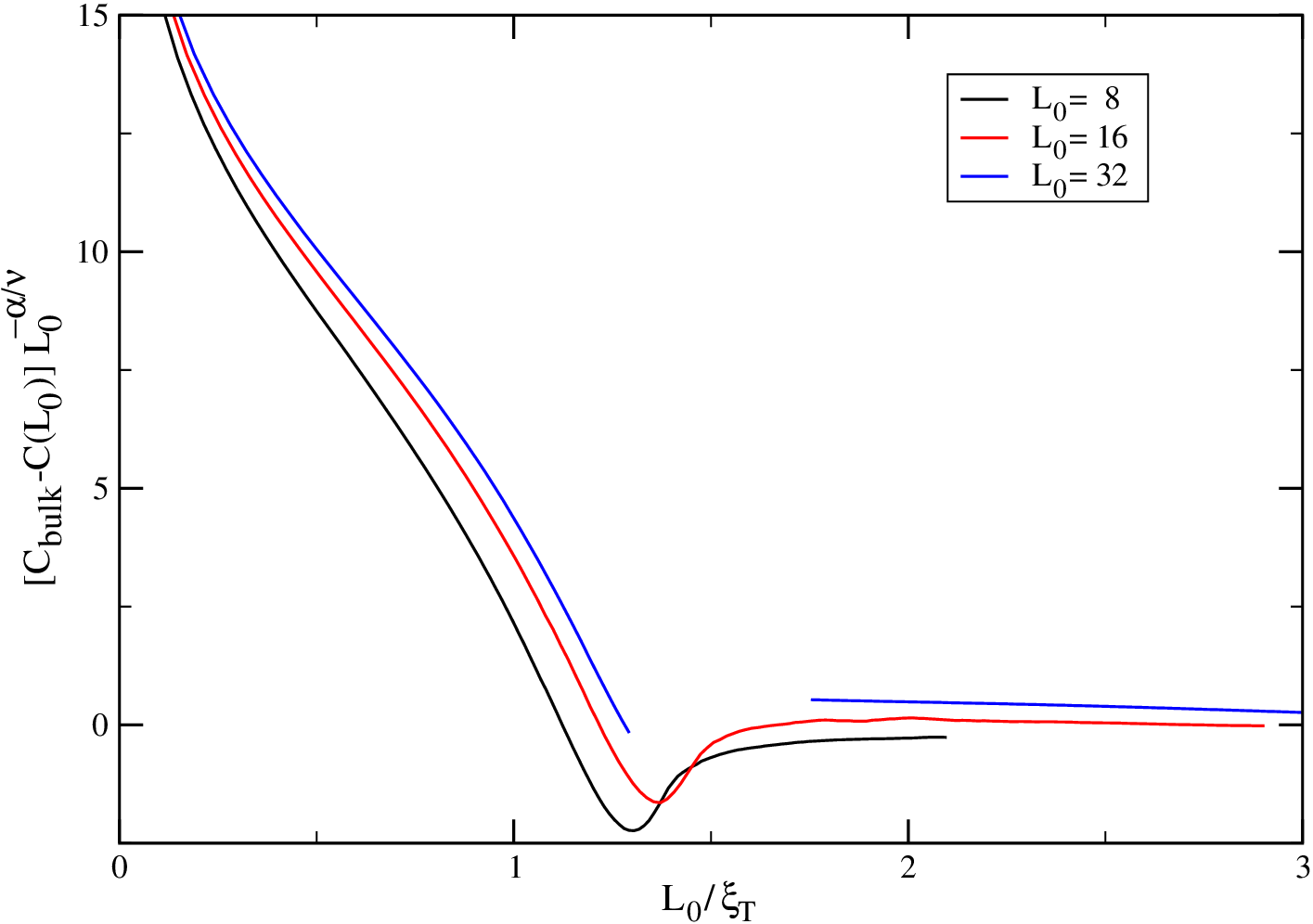}
}
\end{center}
\caption{
\label{F2N}  We plot 
$[C_{bulk}(\beta)-C(\beta,L_0)] L_0^{-\alpha/\nu}$ as a function of
$L_0/\xi_T(\beta)$
for the thicknesses $L_0=8$, $16$ and $32$.  For a discussion see the 
text.
}
\end{figure}

Next we did the same exercise using our data in the low temperature 
phase.
First we compute the finite size scaling function $f_2$ 
without taking into account 
corrections. To this end, in figure \ref{F2N}, we have plotted
$[C_{bulk}(\beta)-C(\beta,L_0) ] L_0^{-\alpha/\nu}$ as a function of 
$L_0/\xi_T(\beta)$, where $\Upsilon(\beta)=1/\xi_T(\beta)$ is given by 
eq.~(\ref{Ups}). As in the high temperature phase,
scaling corrections are clearly visible. Deep 
in the low temperature phase, the function even changes the sign as  
the thickness $L_0$ of the film increases.  Note that these differences
are clearly larger than the statistical errors. As we have
discussed before, the 
error of $[C_{bulk}(\beta)-C(\beta,L_0)]$ should be at most $0.01$ outside of 
the interval $\beta_{max} < \beta < \beta_{KT}$ and maybe up to $0.05$ inside
of this interval. This almost directly translates into the error of
$[C_{bulk}(\beta)-C(\beta,L_0) ] L_0^{-\alpha/\nu}$ since $L_0^{-\alpha/\nu}$
is close to one for $L_0=8,16$ and $32$.

\begin{figure}
\begin{center}
\scalebox{0.62}
{
\includegraphics{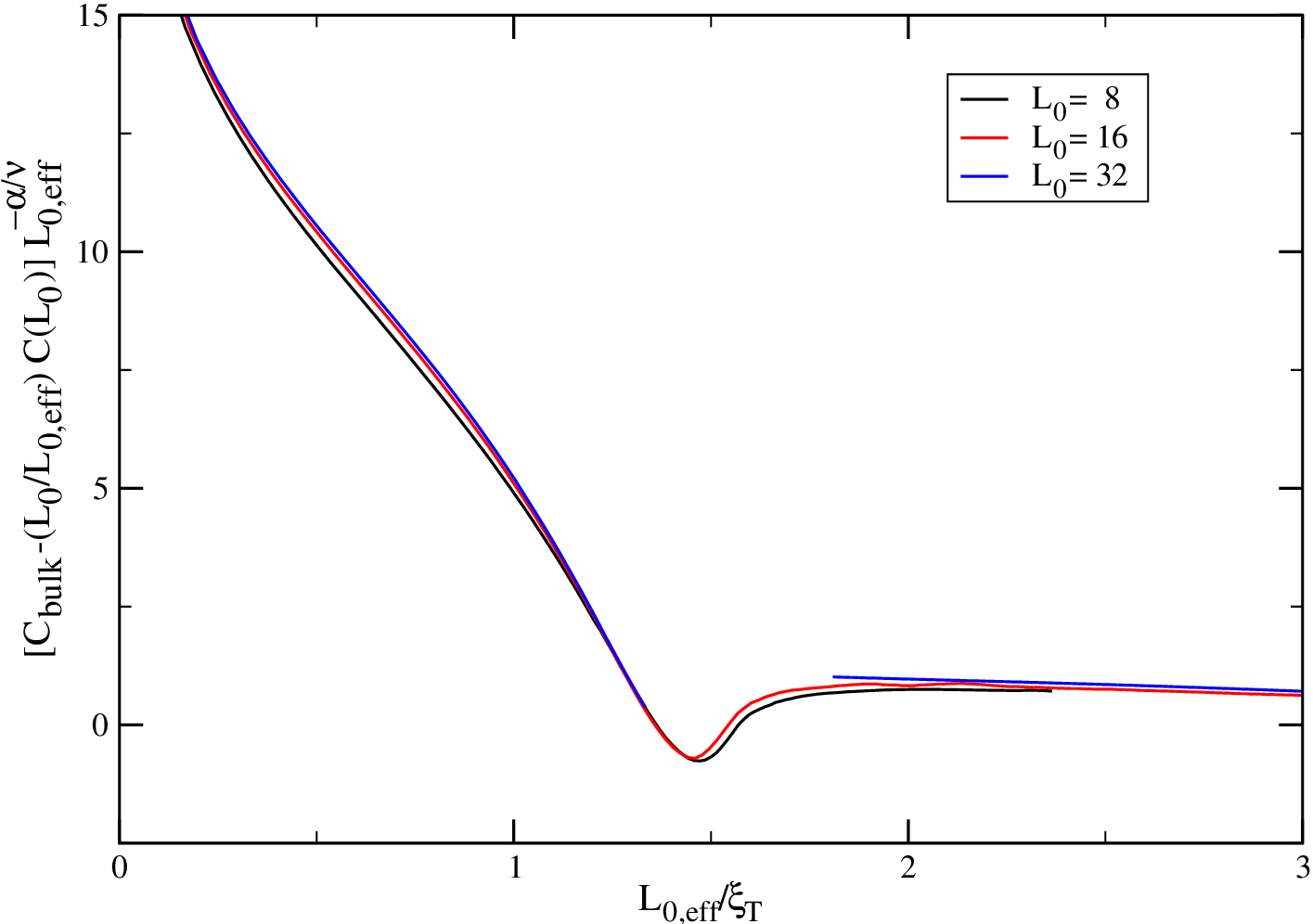}
}
\end{center}
\caption{
\label{F2E}  We plot
$[C_{bulk}(\beta)-C(\beta,L_0) L_0/L_{0,eff}] L_{0,eff}^{-\alpha/\nu}$ 
as a function of $L_{0,eff}/\xi_T(\beta)$
for the thicknesses $L_0=8, 16$ and $32$.  
For a discussion see the text.
}
\end{figure}

In figure \ref{F2E} we have taken into account the leading corrections 
to the singular part of the specific heat by replacing $L_0$ by
$L_{0,eff}=L_0+L_s$.  This has two effects: On the $x$-axis we 
replace $L_0/\xi_T$ by $L_{0,eff}/\xi_T$ and  
the specific heat of the film is multiplied by $L_0/L_{0,eff}$.
After this replacement, the three curves are much closer than in figure 
\ref{F2N}. 
Note that here the error induced by the uncertainty of $L_s=1.02(7)$ 
dominates the error of 
$[C_{bulk}(\beta)-C(\beta,L_0) L_0/L_{0,eff}] L_{0,eff}^{-\alpha/\nu}$.

\begin{figure}
\begin{center}
\scalebox{0.62}
{
\includegraphics{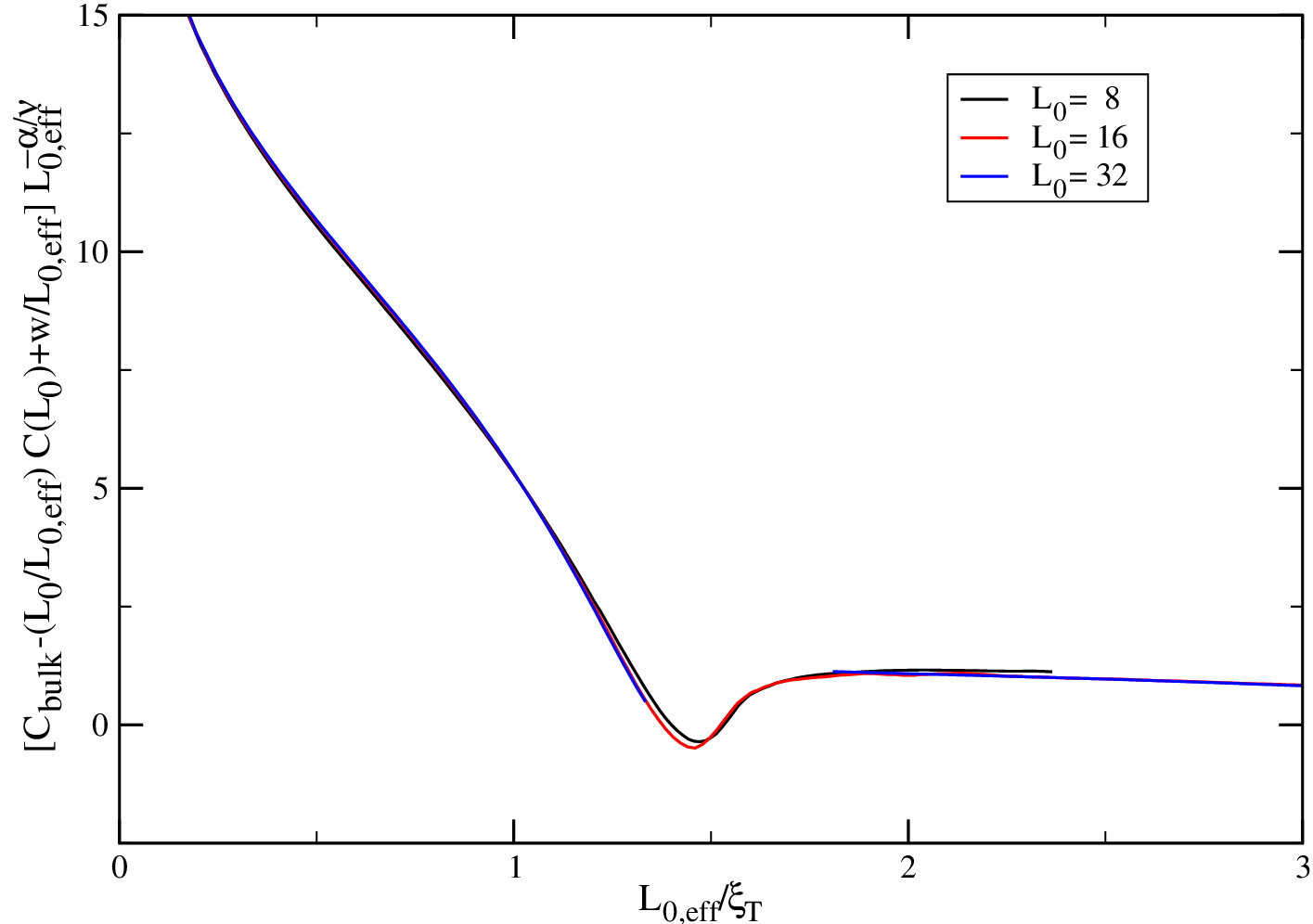}
}
\end{center}
\caption{
\label{F2EW}  We plot
$[C_{bulk}(\beta)-C(\beta,L_0) L_0/L_{0,eff} +w/L_{0,eff}] L_{0,eff}^{-\alpha/\nu}$
with $w=3.5$ as a function of $L_{0,eff}/\xi_T(\beta)$
for the thicknesses $L_0=8, 16$ and $32$.
For a discussion see the text.
}
\end{figure}

Finally, in figure \ref{F2EW} we have taken into account the 
correction $w/L_{0,eff}$, where we have set $w=3.5$.
Now the curves fall nicely on top of each other giving support to the 
suggestion made in section \ref{theory} that boundary correction to the 
analytic background are not well described by replacing $L_0$ 
by $L_{0,eff}=L_0+L_s$. Note that now the uncertainty of $w$ gives 
the largest contribution to the error of 
$[C_{bulk}(\beta)-C(\beta,L_0) L_0/L_{0,eff} +w/L_{0,eff}] 
L_{0,eff}^{-\alpha/\nu}$.

In tables \ref{f2number32} and \ref{f2number16} 
we give $f_2$ for a few values of $L_{0,eff}/\xi_{T}$.
This should help the reader to compare our result with that 
obtained from other systems.

\begin{table}
\caption{\sl \label{f2number32}
We give $f_2$ for a few 
values of $L_{0,eff}/\xi_{T}$ (upper row). These numbers are 
obtained from $L_0=32$ and corrections characterized by
$L_s=1.02$ and $w=3.5$.  Assuming $L_s=1.02$ and $w=0$, the values
of  $f_2$ are lower by about $0.11$ throughout. 
}
\begin{center}
\begin{tabular}{|ccccccccccccc|}
\hline
0.4   & 0.5 & 0.6  & 0.7  & 0.8  & 0.9  & 1.0 & 1.1  & 1.2 & 1.3 & 1.9 & 2.0 &2.5 \\
11.75 &10.67&9.66  & 8.67 & 7.64 & 6.54 & 5.33& 3.98 & 2.50&0.96 & 1.11& 1.08
& 0.97\\
\hline
\end{tabular}
\end{center}
\end{table}

\begin{table}
\caption{\sl \label{f2number16}
We give $f_2$ for a few
values of $L_{0,eff}/\xi_{T}$ (upper row).  These
results are obtained from $L_0=16$ and corrections characterized by
$L_s=1.02$ and $w=3.5$.  Assuming $L_s=1.02$ and $w=0$, the values
of $f_2$ are lower by about $0.22$ throughout. 
}
\begin{center}
\begin{tabular}{|ccccc|}
\hline
1.4    &  1.5   &  1.6 &  1.7  &  1.8 \\
-0.22  & -0.25  &  0.67&  0.94 &  1.02 \\
\hline
\end{tabular}
\end{center}
\end{table}

The finite size scaling function shows a clear minimum at a finite 
value of $[L_{0,eff}/\xi_T]$. By construction, 
the position of this minimum does not depend on $w$. For
$L_0=16$ we get $[L_{0,eff}/\xi_T]_{min}  \approx 1.452$. For $w=3.5$
the minimum takes the value $-0.47$ and for $w=0$ the value $-0.69$.
For $L_0=8$ we get $[L_{0,eff}/\xi_T]_{min}  \approx 1.468$ and as
value $-0.36$ for $w=3.5$ and $-0.76$ for $w=0.0$.  We consider it as 
a robust result that $f_2$ assumes a negative value at its minimum.
Note that the KT transition takes place at $L_0/\xi_T = 1.595(7)$
\cite{myKTfilm}. I.e. the minimum of $f_2$ is located at a temperature
slightly higher than the transition temperature.

\begin{figure}
\begin{center}
\scalebox{0.62}
{
\includegraphics{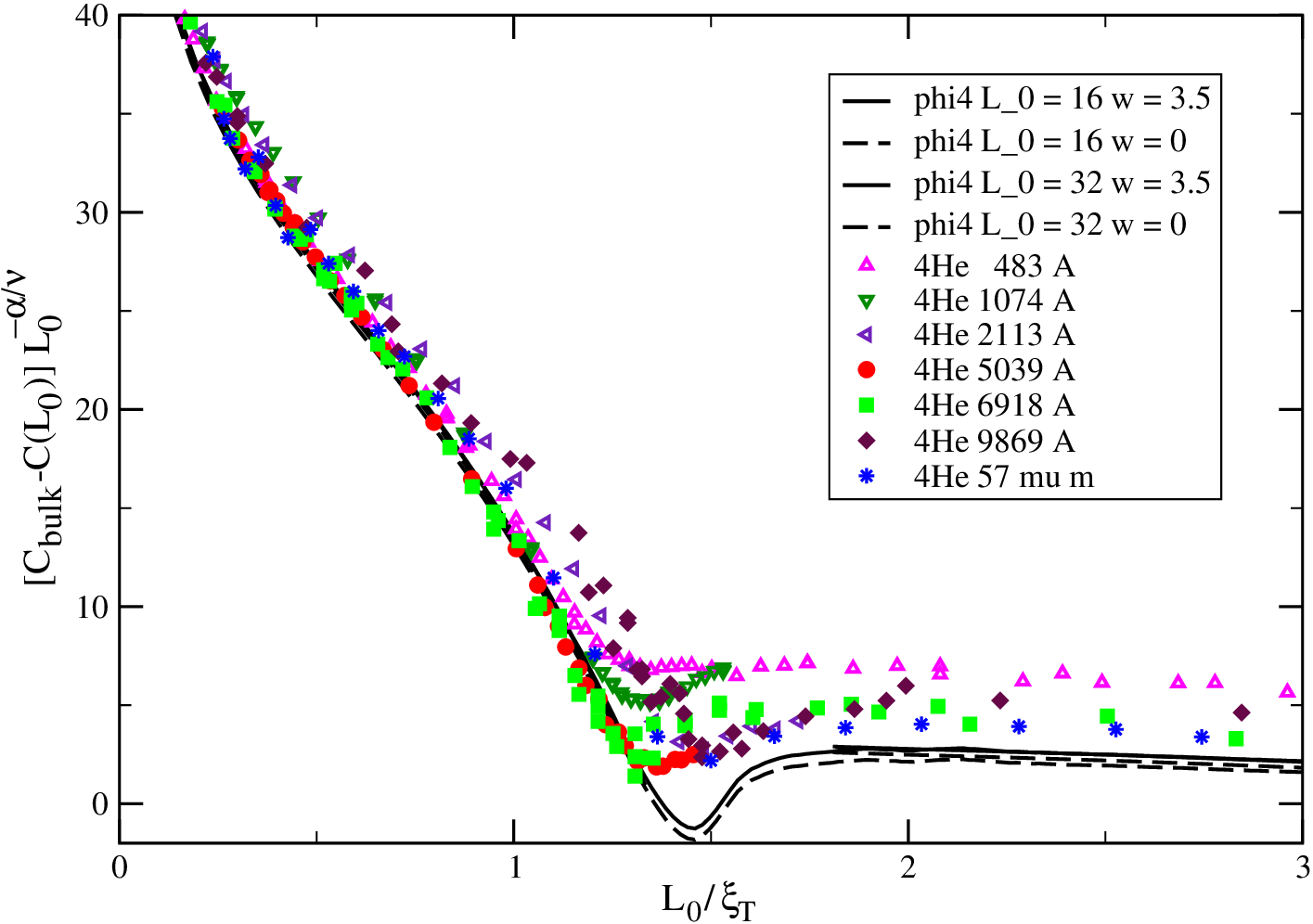}
}
\end{center}
\caption{
\label{F2EXPL}  
Experimental results
\cite{KiMeGa99,KiMeGa00,Lietal00} obtained for films
of $^4$He in the low temperature phase, close to the $\lambda$-transition.
We plot $[C_{bulk}(t)-C(t,L_0)] L_{0}^{-\alpha/\nu}$ as a function of
$L_0/\xi_{T}(t)$ for the thicknesses $L_0=483$,
$1074$, $2113$, $5039$, $6918$, $9869 \AA$ and $57$ $\mu$m.
For comparison we give our results obtained for $L_0=16$ and $32$. 
In contrast to
the experimental results, boundary corrections, characterized by $L_s=1.02$
and $w=3.5$ (solid black lines) or by $L_s=1.02$ and $w=0$ (dashed black lines)
are taken into account. Our results for
$f_2$ are multiplied by $r_{^4He,\phi^4}=2.57$.
For a discussion see the text.
}
\end{figure}

In figure \ref{F2EXPL} we compare our result  for $f_2$
with experimental ones.
To this end, we plot  $[C_{bulk}(t)-C(t,L_0)] L_0^{-\alpha/\nu}$
as a function of $L_0/\xi_{T}(t)$ using the experimental data of 
\cite{KiMeGa99,KiMeGa00}  for thin films of $^4$He at vapor pressure
of the thicknesses
483, 1074, 2113, 5039, 6918  and $9869 \AA$. These data
are taken from the web page \cite{Gaspariniweb}.
The experimental data for the specific heat are given as a function of the
reduced temperature $t=(T-T_{\lambda})/T_{\lambda}$. In order to plot
them as a function of $L_0/\xi_{T}(t)$ we use
$\xi_{T}=3.45 \; |t|^{-\nu} \AA$, eq.~(\ref{xi0T}). As value of the
critical exponents we take
$\nu=0.6717$ \cite{recentXY} and correspondingly $\alpha=-0.0151$.
Furthermore we plot the results of \cite{Lietal00} for a film of 57 $\mu$m
as thickness. We have rescaled these data as discussed in the previous
subsection on the high temperature phase.
For comparison we give
our result for $f_2$ obtained from $L_0=16$ and $32$,
taking into account boundary
corrections characterized by $L_s=1.02$ and $w=3.5$ (solid black lines).
To indicate the
possible errors we give in addition the results
obtained for $L_s=1.02$ and $w=0$ (dashed black lines).
We have multiplied our numbers by $r_{^4He,\phi^4}=2.57$, eq.~(\ref{fuckor}).

Here almost all experimental results for the finite size scaling 
function $f_2$ are somewhat larger than ours. 
There is some scattering among the experimental results.  
Those for the larger thicknesses of the film are
closest to our $f_2$. There is a nice match for the position of the 
minimum of $f_2$ obtained from $57$ $\mu$m  and 
our result.  However  $f_2$, computed by using the experimental data,
never assumes negative values. Also the dip around the minimum is much less 
pronounced than it is in our case.
It is beyond the scope of the present work to discuss possible sources of
these discrepancies. This requires detailed discussions with the 
experimentalists.

\subsection{The finite size scaling function $f_2$ in the neighbourhood of the
critical point}
Here we like to give an explicit formula for $f_2$ in the neighbourhood 
of $t=0$. To this end, we use the results for the amplitudes $A_{\pm}$ 
obtained in subsection \ref{threeD}, the amplitudes of the correlation 
length $\xi_{0,2nd}$ and $\xi_{0,T}$ and the value of $f_2(0)$ obtained 
in subsection \ref{FSSTC}. In addition we have extracted from our 
data the slope of the specific heat of thin films at the transition 
temperature of the three-dimensional bulk system.

As result we get in the high temperature phase:
\begin{equation}
f_2=161.6(6) - 2.1 \delta \;
- [158.3(6) -2 \delta ](L_0/\xi_{2nd})^{-\alpha/\nu} 
+ 0.65(2) (L_0/\xi_{2nd})^{1/\nu} + \ldots
\end{equation}
where $\delta=1/\alpha+1/0.0151$ gives the dependence on the value of 
$\alpha$ that is used for the analysis. This  formula
gives a good approximation of 
$f_2$ up to about $L_0/\xi_{2nd}=0.7$, where the deviation from the full 
result as computed above is less than 1$\%$.

In the low temperature phase we obtain:
\begin{equation}
f_2=161.6(6) - 2.1 \delta \;
- [152.2(6) -2 \delta ](L_0/\xi_{T})^{-\alpha/\nu}
- 2.43(5) (L_0/\xi_{T})^{1/\nu} + \; \ldots
\end{equation}
This is a good approximation of
$f_2$ up to about $L_0/\xi_{T}=0.3$, where the deviation from the full
result as computed above is less than 1$\%$.

\subsection{The finite size scaling function $f_2$ for large $|t|$}
\label{larget}
In the high temperature phase, for $L_0 \gg \xi_{2nd,3D}$, the two boundaries
are uncorrelated and therefore the dependence of physical quantities on 
$L_0$ is trivial.  In the case of the specific heat we can write 
(ignoring boundary corrections)
\begin{equation}
L_0 C(\beta,L_0) = L_0 C_{bulk}(\beta) + 2 C_s(\beta) \;\;\; \mbox{for} 
\;\; \beta < \beta_c \;\;, \; L_0 \gg \xi_{2nd,3D} \;\;.
\end{equation}
Inserting this equation in the definition of the scaling function $f_2$ we
get 
\begin{equation}
\label{f2larget}
 f_2 = -2 C_s(\beta) L_0^{-1-\alpha/\nu}
 \;\;\; \mbox{for} \;\; L_0 \gg \xi_{2nd,3D}
\end{equation}
Since $f_2$ is a function of $L_0/\xi_{2nd}$ only and $C_s(\beta)$ does not 
depend on $L_0$, it follows
\begin{equation}
 C_s = h_+ \xi_{2nd}^{1+\alpha/\nu}
\end{equation}
following the convention in the literature 
\begin{equation}
h_+ = \frac{A_s^+  \xi_{0,2nd}^{-1-\alpha/\nu} }{\alpha + \nu}
\end{equation}
In the low temperature phase, the situation is more complicated,  due
to the presence of the Goldstone mode.  Power like corrections are 
present in this case:
\begin{equation}
L_0 C(\beta,L_0) = L_0 [C_{bulk}(\beta) +\mbox{O}(L_0^{-3})] + 2
C_s(\beta) \;\;\; \mbox{for}  \;\; \beta > \beta_c \;\;, \;
 L_0 \gg \xi_{T,3D} \;\;.
\end{equation}
We do not have data for a sufficiently large range of temperatures to 
check carefully this behaviour. 
Assuming the correctness of eq.~(\ref{f2larget}) we read off from 
our data for $L_0=32$:
\begin{equation}
 h_+=-2.2(2)
 \end{equation}
in the high temperature phase and
\begin{equation}
 h_- =-1.2(2)   \;\;,
\end{equation}
in the low temperature phase, where
\begin{equation}
h_- = \frac{A_s^-  \xi_{0,T}^{-1-\alpha/\nu} }{\alpha + \nu}
\end{equation}
In both case, we have taken into account the boundary corrections. 
It follows that
$A_s^+ = - 5.2(5)$ and  $A_s^- = -6.8(1.1)$   in units of 
$J \AA \mbox{mole}^{-1} K^{-1}$.  
Furthermore, we can compute the universal ratio
\begin{equation}
 Q= \frac{A_s^{+} }{A_s^{-} } = 0.8(2)  \;\;.
\end{equation}
Experimental results for films of $^4$He have been summarized  by the 
authors of \cite{GaKiMoDi08}  as
\begin{equation}
 A_s^+  = - 5.9 \pm 0.2   \;\;\;,\;\; A_s^-  = - 8.6 \pm 0.5
\end{equation}
in units of $J \AA \mbox{mole}^{-1} K^{-1}$.
It follows 
\begin{equation}
 Q = \frac{A_s^{+} }{A_s^{-} } = 0.69(5)   \;\;.
\end{equation}
Our result for $A_s^+$ is consistent with that of experiments. In the 
case of $A_s^-$ we see a small discrepancy. The results for $Q$ are 
consistent within the error bars.

The surface specific heat has been calculated using the perturbative
expansion in three dimensions fixed in the two-loop approximation by 
Mohr and Dohm
\cite{MoDo00,Mohr00}. Inserting numerical values into eqs.~(3,4)  
one gets
$A_s^+ = - 5.429$ and $A_s^-=- 1.822$ in units of 
$J \AA \mbox{mole}^{-1} K^{-1}$
\cite{Dohmprivat09}.
We notice that the value for $A_s^+$ is in excellent agreement with
our results and also close to the experimental one. In contrast, the 
value for $A_s^-$ is clearly ruled out by us as well as by the experiment.
Already the authors of \cite{MoDo00} have pointed out that their result
for $A_s^-$ does not provide an accurate numerical estimate.  

Krech and Dietrich quote as result of the $\epsilon$-expansion, 
eq.~(E6) of  \cite{KrDi92}:
\begin{equation}
 A_{s}^+ \xi_{0,+}^{d-1} = - \frac{N}{256 \pi} 
 \left\{2 + \epsilon \left[2 + 
 \ln \pi -\gamma + \frac{N+2}{N+8} \right] +\mbox{O}(\epsilon^2) \right\} \;.
\end{equation}
Inserting $N=2$ one gets
\begin{equation}
A_{s}^+ \xi_{0,+}^{d-1} =- 0.0049736 - 0.0073796 \epsilon + \ldots
\end{equation}
In order to compare with the experiments on superfluid $^4$He at 
vapour pressure we insert $\xi_0^+=1.422 \AA$ and then convert from units
of $\AA^{-2}$ to $J \AA/ \mbox{mole}^{-1} K^{-1}$.  We arrive at
\begin{equation}
A_{s}^+ =[- 1.88  - 2.79 \epsilon + \ldots]  J \AA/ \mbox{mole}^{-1} K^{-1} \;\;
\end{equation}

Following Eisenriegler \cite{Ei84}  
\begin{equation}
\frac{A_{s}^+}{A_{s}^-} = \frac{\pi}{2^{3/2}} \frac{N}{N+8} \epsilon 
+ \mbox{O}(\epsilon^2) = 0.222  \epsilon + \mbox{O}(\epsilon^2)
\end{equation}
where we have inserted $N=2$.
The result of Eisenriegler has been extended by Upton \cite{Up98} 
to $\mbox{O}(\epsilon^2)$.
For $N=2$ one gets:
\begin{equation}
\frac{A_{s}^+}{A_{s}^-} =  0.222  \epsilon [1 + 0.93 \epsilon + \ldots]
\end{equation}
For a detailed discussion of the field theoretic results see the 
PhD thesis of Mohr \cite{Mohr00}.

\section{The finite size scaling function $f_1$}
In order to compute the finite size scaling function $f_1$ we have to determine
the specific heat at the temperatures, where $\xi_{2nd}$ of the 
three-dimensional bulk system assumes the values 
$8$, $16$ and $32$ or,  taking into account corrections,
$9.02$, $17.02$ and $33.02$.   To this end, we have first numerically 
inverted eq.~(\ref{xires}).   The corresponding values of $\beta$ are
$0.50295$, $0.50694$, $0.50836$ and $0.50396$, $0.50713$, $0.50840$,
respectively.  At these values of $\beta$ the specific heat of the 
three-dimensional bulk system assumes the values
$7.091(2)$,  $9.277(2)$,  $11.512(4)$  and 
$7.462(2)$,  $9.475(2)$,  $11.613(4)$, 
respectively. These values are obtained by using the results of fits
with the ansatz~(\ref{critical}).
The error is dominated by the uncertainty of $C_{ns}$ that has been used 
as input in eq.~(\ref{critical}).

\begin{figure}
\begin{center}
\scalebox{0.62}
{
\includegraphics{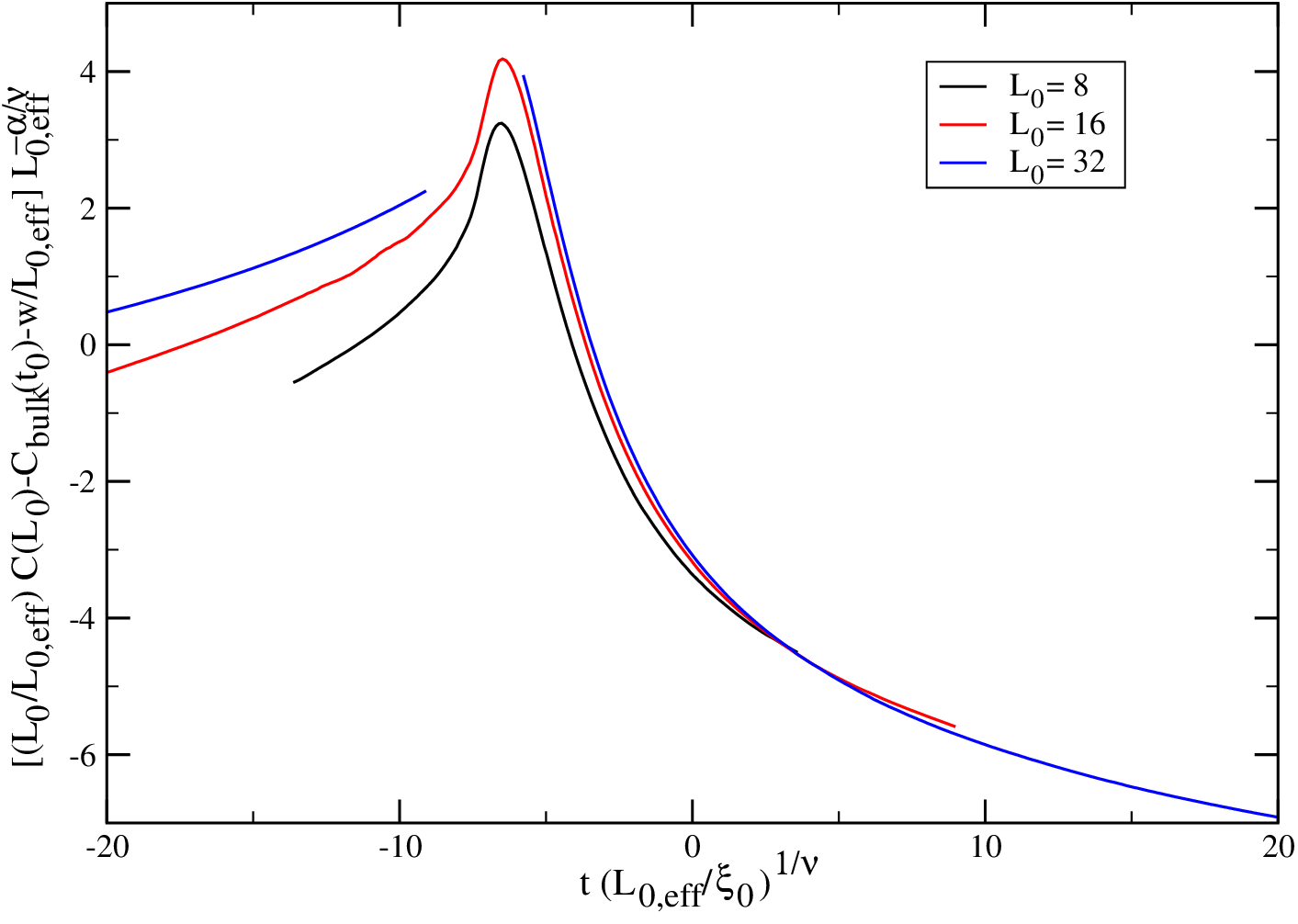}
}
\end{center}
\caption{
\label{F1EW}  We plot
$[\frac{L_0}{L_{0,eff}} C(t,L_0)-C_{bulk}(t_0)-w/L_{0,eff}] 
L_{0,eff}^{-\alpha/\nu}$
versus $t (L_{0,eff}/\xi_{0,2nd})^{1/\nu}$, where we use $L_s=1.02$ and
$w=3.5$.
For a discussion see the text.
}
\end{figure}

\begin{figure}
\begin{center}
\scalebox{0.62}
{
\includegraphics{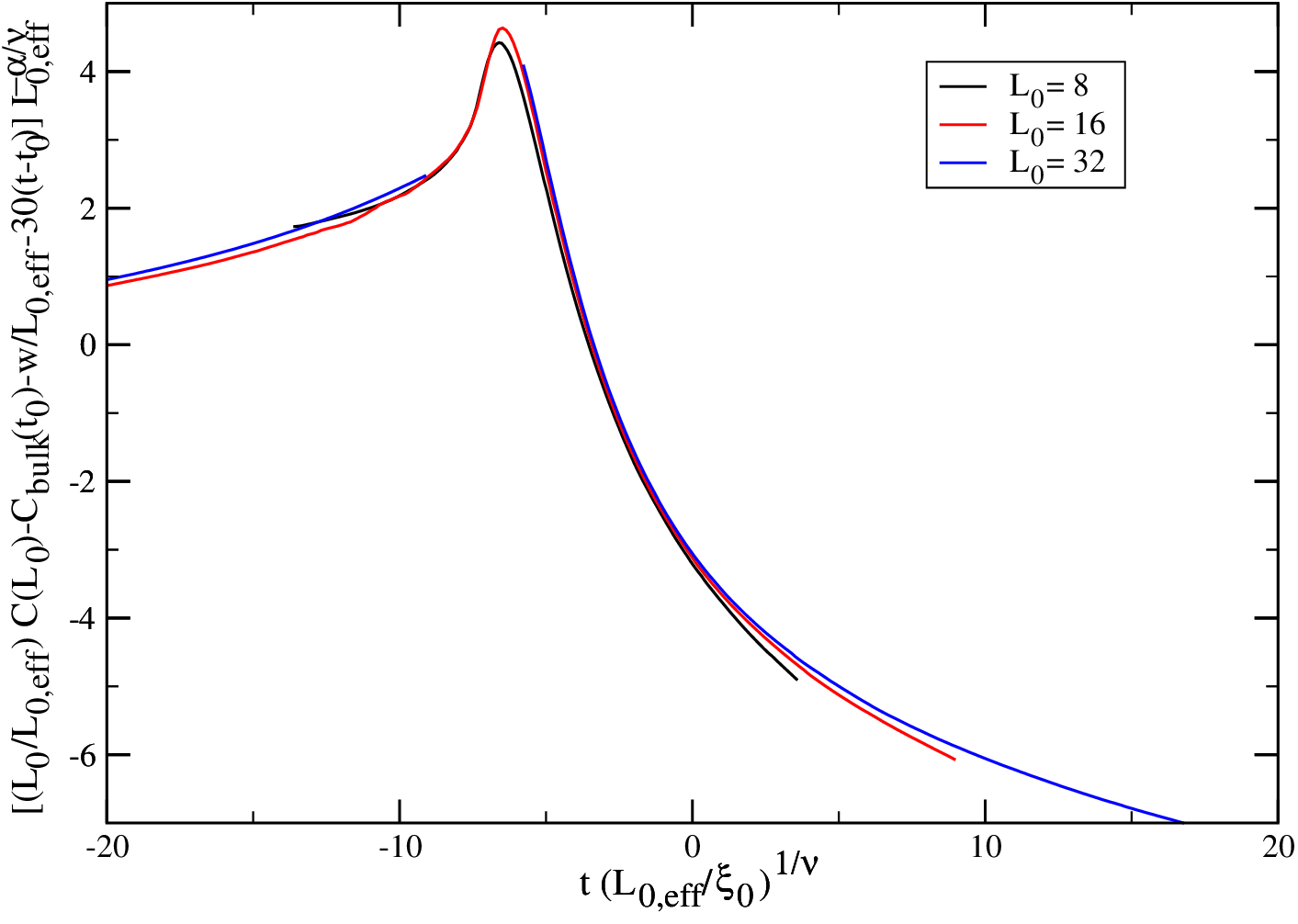}
}
\end{center}
\caption{
\label{F1EWX}  We plot
$[\frac{L_0}{L_{0,eff}} C(t,L_0)-C_{bulk}(t_0)-30(\beta_0-\beta)] 
L_{0,eff}^{-\alpha/\nu}$
as a function of $t (L_{0,eff}/\xi_{0})^{1/\nu}$
For a discussion see the text.
}
\end{figure}

In figure \ref{F1EW} we plot 
$[\frac{L_0}{L_{0,eff}} C(t,L_0)-C_{bulk}(t_0)-w/L_{0,eff}] L_{0,eff}^{-\alpha/\nu}$
as a function of $t (L_{0,eff}/\xi_{0,2nd})^{1/\nu}$, 
where we have used $L_s=1.02$ and $w=3.5$. In particular
for the low temperature phase we see a strong dependence on $L_0$. 
We argue that this is due to analytic corrections.  In the case of
$f_2$ these corrections affect the specific heat of the  
three-dimensional bulk system and the thin film in the same way. Therefore 
there is only a rather small effect on the difference of the two. 
In the case of $f_1$  the specific heat of  the three-dimensional
bulk system and the thin film are taken at different reduced 
temperatures. Therefore there is a quite huge effect on the difference 
of the two.
Motivated by this, in figure \ref{F1EWX}, we have subtracted 
$C_{bulk}(t_0)+ 30 \times (\beta_0-\beta)$ instead of $C_{bulk}(t_0)$.
The coefficient $30$ is fitted by eye to get a reasonable 
collapse of the curves at low temperatures.

\begin{figure}
\begin{center}
\scalebox{0.62}
{
\includegraphics{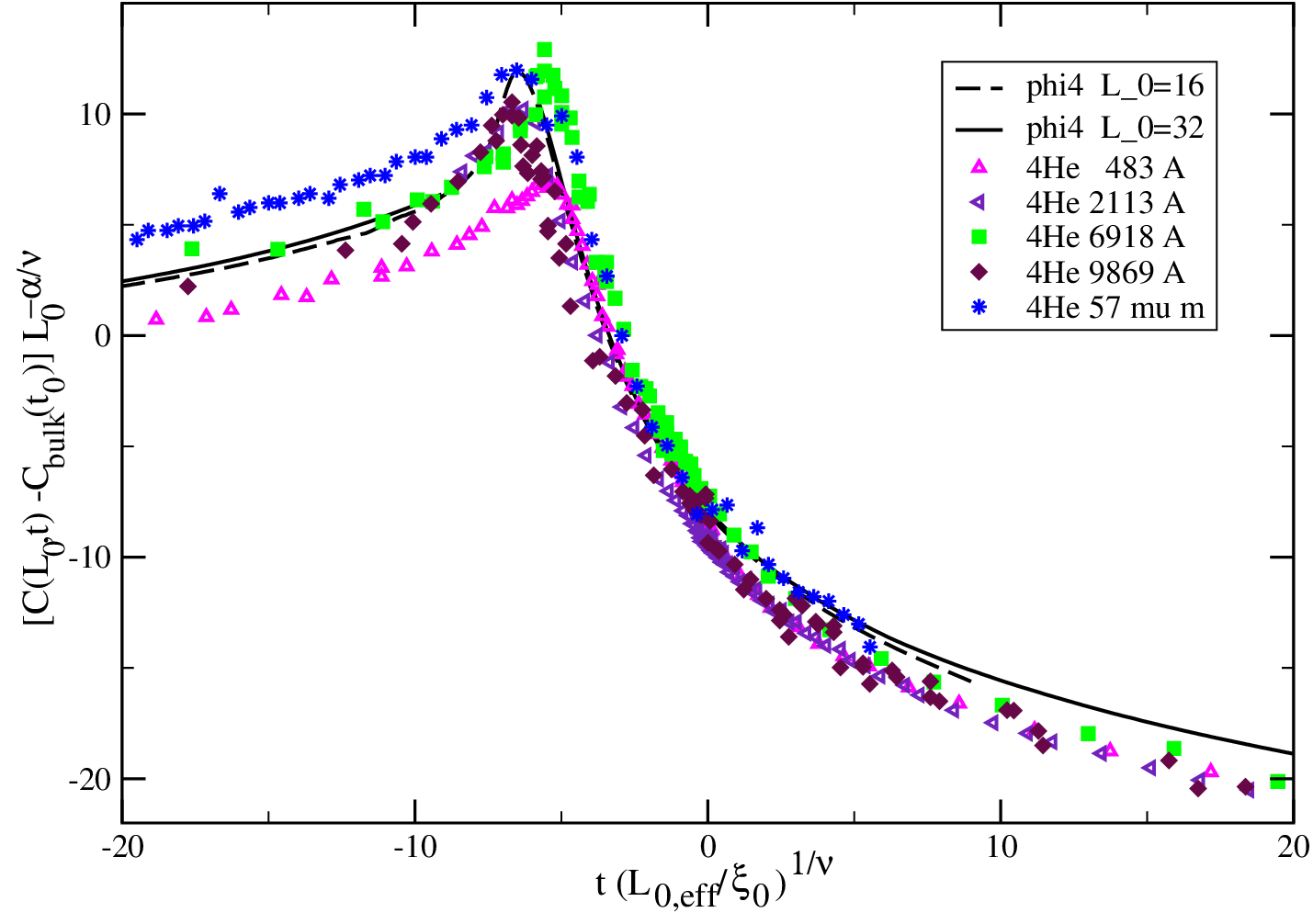}
}
\end{center}
\caption{
\label{F1EXP}  In the figure we plot
$[C(t,L_0)-C_{bulk}(t_0)] L_{0}^{-\alpha/\nu}$
as a function of $t (L_{0}/\xi_{0})^{1/\nu}$ for experimental data.
For comparison we give our results of figure \ref{F1EWX} for $L_0=16$ and $32$.
our numbers have been multiplied by  $r_{^4He,\phi 4}=2.57$.
For a discussion see the text.
}
\end{figure}

In figure  \ref{F1EXP} we compare with experimental results for the 
finite size scaling $f_1$. In order to compute $t_0$ we have 
used $\xi_{2nd}=1.422 \AA t^{-0.6717}$, eq.~(\ref{xi02nd}). 
We have plotted $[C(t,L_0)-C_{bulk}(t_0)] L_0^{-\alpha/\nu}$ as a function 
of $t (L_0/\xi_{0,2nd})^{1/\nu}$, where the specific heat is given in units of
$J \mbox{mole}^{-1} K^{-1}$ and $L_0$ in $\AA$. 
To this end we have used the data of
\cite{KiMeGa99,KiMeGa00}  for thin films of $^4$He at vapor pressure
of the thicknesses
483, 2113, 6918  and $9869 \AA$. These data
are taken from the web page \cite{Gaspariniweb}. For a better readability 
of the figure we do not give the data for $1074$ and  $5039 \AA$.
In addition we give the results of \cite{Lietal00}. In this case, we have
computed $C_{bulk}(t_0)$ from the results of the fits given in 
ref. \cite{lipa2003}.    
For comparison we have taken our results for $L_0=16$ and $32$ 
from figure \ref{F1EWX}.
In order to match with the experimental results, 
we have multiplied our numbers by $r_{^4He,\phi 4} = 2.57$.  
In the high temperature phase, the experimental results fall nicely 
on top of each other. In contrast, in the low temperature phase, 
in particular for temperatures below the position of the peak, we see some
scattering of the experimental results.
In the main our results are compatible with those of the experiments
on $^4$He films. In the high temperature phase our result is
slightly larger than the experimental one.

Finally let us compare our results with those obtained from field 
theory and from previous Monte Carlo simulations.
In figure 1 (a) of \cite{ScWaDoFr90} a one-loop result for the
finite size scaling  function $f_1$ is given.  The specific heat is
given in units of $J \mbox{mole}^{-1} K^{-1}$ and the length in units of $\AA$.
The function has similar
qualitative features as our result. However the peak in the low
temperature phase is much more shallow than in our case. The
maximal value is about 7,  while we get 11.9. The position of the
peak ($t L_0^{1/\nu}\approx 9$) slightly differs from ours
($t L_0^{1/\nu} \approx 10.9$).
In  \cite{ScMa95} the function $f_1$ has been calculated from
Monte Carlo simulations of the standard XY model in three dimensions.
The authors have used staggered boundary conditions in order to
suppress the order parameter at the boundary.  
They have simulated lattices of the
thicknesses $L_0 \leq 24$ and $L_1=L_2 \leq 100$.
They define the specific heat as the second
derivative of the free energy with respect to the temperature. They compute
it using
\begin{equation}
C=\frac{\beta^2}{L_0 L_1 L_2} \left[ \langle E^2 \rangle -
 \langle E \rangle^2  \right] \;\;.
 \end{equation}
Their final result is 
given in figure 4 of \cite{ScMa95}.  See also, e.g., figure 2 of 
\cite{Lietal00}.
From the scattering of the data, we conclude that statistical errors are 
much larger than in our case. Therefore the authors were not able to conduct
a detailed analysis of corrections to scaling as we did here.
Their result looks qualitatively the same as ours. The value of the peak seems
to be somewhat larger than in our case.  Also the position of the peak is 
slightly different; from figure 4 of  \cite{ScMa95} we read off   
$t L_0^{1/\nu} \approx -9$.
The authors of \cite{NhMa03} have simulated the standard XY model using 
films of the thicknesses $L_0=12,14$ and $16$.  Throughout, they have used
$L_1=L_2=5 L_0$.  They have used free (``open'' in their notation)
boundary conditions in the short direction.
 Their result for $f_1$ is presented in figure 2 of \cite{NhMa03}.
From the scattering of the 
results it is clear that statistical errors are much larger than in our case.
The results of \cite{ScMa95} and \cite{NhMa03} are consistent.
There is also reasonable agreement with experimental results.
Similar to \cite{ScMa95} the value of the maximum seems to be larger than 
in our case, also the position of the peak is slightly different 
($t L_0^{1/\nu} \approx - 8.5$). Also in \cite{NhMa03}
corrections to scaling are not discussed.

\section{Summary and Conclusions}
We have studied the finite size scaling behaviour of the 
specific heat of thin films in the three-dimensional XY universality
class. To this end 
we have simulated the improved two-component $\phi^4$ model on the 
simple cubic lattice.  In order the get a vanishing order parameter
at the boundary, which is observed in experiments on films of $^4$He 
near the $\lambda$-transition, we have employed free boundary conditions.
These can be interpreted as Dirichlet boundary conditions with 
the value $\vec{\phi}=(0,0)$ of the field at the boundary.
We discuss how leading boundary corrections affect the finite size 
scaling behaviour of the specific heat of thin films. 
We point out that the analytic part of the specific heat might suffer from 
boundary corrections that are not described by $L_{0,eff}=L_0+L_s$, which 
characterizes the leading corrections to the singular part.

First we have performed simulations to get the energy density of the 
three-dimensional system for a large number of temperatures. 
These simulations supplement those of  \cite{myAPAM,myamplitude}. 
Using these data we computed accurate estimates of the specific heat 
in the range $0.49 < \beta < 0.58$  of inverse temperatures.

Next we have analysed in detail the finite size scaling behaviour of the 
specific heat of thin films at the $\lambda$-transition. To this 
end we have simulated films up to a thickness of $L_0=64$ lattice units.
Our result is in nice agreement with that obtained for thin films of 
$^4$He at the $\lambda$-transition \cite{GaKiMoDi08}. 

Furthermore we have simulated films of the thicknesses 
$L_0=8,16$ and $32$ for a large range of inverse temperatures $\beta$ in the 
neighbourhood of the $\lambda$-transition.
We have taken great care to obtain reliable estimates for the 
two-dimensional thermodynamic limit of the thin films. Using our data
we have computed the finite size scaling functions $f_1$ and $f_2$ defined 
in the introduction. It turns out that corrections to scaling which are 
caused by the free boundary conditions, have to be taken into account, to 
get a good collapse of the data obtained for different thicknesses of the 
film. These corrections can be described,
to leading approximation, by an effective thickness 
$L_{0,eff} = L_0 +  L_s$ of the film.  In  \cite{myKTfilm} we have 
obtained $L_s=1.02(7)$ from a finite size scaling study at the critical
point of the three-dimensional system. However, a priori, this $L_{0,eff}$
only applies to the singular part of the specific heat. Our analysis of 
the data shows that in fact the analytic part of the specific heat 
requires an additional correction which is $\propto L_0^{-1}$.

The comparison of our results for the finite size scaling functions 
$f_1$ and $f_2$ and those obtained 
from experiments on films of $^4$He at the $\lambda$-transition in
general show nice agreement. We think that, 
in order to explain the minor discrepancies,  
a detailed knowledge of the experimental work is required. Therefore 
we abstain from any speculation on the sources of these discrepancies.

We have also compared with results obtained from field theory and previous 
Monte Carlo simulations. Field theoretic calculations are of low order;
O($\epsilon$) in the case of the $\epsilon$-expansion and one or two-loop
in the case of the perturbative expansion in three dimensions fixed.
Previous Monte Carlo simulations are effected by relatively large 
statistical errors. Corrections to finite size scaling were not discussed
in these works.

\section{Acknowledgements}
This work was supported by the DFG under grants No JA 483/23-1 and 
HA 3150/2-1. The simulations
were performed on the compute cluster GRAWP at the Institute for 
theoretical physics of the Universit\"at Leipzig and on various computers
at the physics department of the Humboldt-Universit\"at zu Berlin.
I like to thank  W. Janke for discussions and support, V. Dohm, M. Kimball 
and J. Lipa for sending data and helping me with the literature.

\end{document}